\documentclass{article}
\usepackage{amssymb}
\usepackage{amsmath}
\usepackage{amsfonts,amsthm} 
\usepackage{titlesec}
\usepackage{graphicx}
\usepackage[hidelinks]{hyperref}
\usepackage{float}
\usepackage{mathtools}
\usepackage{afterpage}
\usepackage[export]{adjustbox}
\usepackage{graphicx}
\usepackage{subcaption}
\usepackage{dsfont}
\usepackage{enumitem}
\usepackage{algorithm}
\usepackage{algpseudocode}

\DeclarePairedDelimiter{\ceil}{\lceil}{\rceil}
\usepackage{color}
\usepackage[english]{babel}
\titleformat{\chapter}[display]
  {\normalfont\LARGE\bfseries}
  {\titleline{}\vspace{5pt}\titleline{}\vspace{1pt}%
  \MakeUppercase{\chaptertitlename} \thechapter}
  {1pc}
  {\titleline{}\vspace{0.5pc}} 
\usepackage{mathtools}
\DeclarePairedDelimiter\abs{\lvert}{\rvert}

\DeclareMathOperator*{\argmin}{arg\,min}

\makeatletter
\renewcommand\section{\@startsection {section}{1}{\z@}%
                               {-3.5ex \@plus -1ex \@minus -.2ex}%
                               {2.3ex \@plus.2ex}%
                               {\normalfont\large\bfseries}}
\renewcommand\subsection{\@startsection{subsection}{2}{\z@}%
                                 {-3.25ex\@plus -1ex \@minus -.2ex}%
                                 {1.5ex \@plus .2ex}%
                                 {\normalfont\bfseries}}

\makeatother

\usepackage[english]{babel} 

\usepackage{bold-extra}

\usepackage{fancyhdr}        
\numberwithin{equation}{section}       
\numberwithin{figure}{section}         
\numberwithin{table}{section}          

\newtheorem{theorem}{Theorem}[section]
\newtheorem{prop}[theorem]{Proposition}
\newtheorem{corollary}[theorem]{Corollary}
\newtheorem{lemma}[theorem]{Lemma}
\theoremstyle{remark}
\newtheorem{remark}{Remark}[section]
\theoremstyle{definition}

\newtheorem{assumption}{Assumption}
\theoremstyle{definition}

\graphicspath{{img/}}

\title{
\normalfont \normalsize   
\huge On theoretical guarantees and a blessing of dimensionality for nonconvex sampling      
}
\author{Martin Chak}
\date{\normalsize\today}

\begin{document}
\maketitle

\begin{abstract}
Guarantees for algorithms sampling from nonlogconcave target measures on~$\mathbb{R}^d$ are studied. For the class of measures with logdensities that have bounded Hessians and are strongly concave outside a Euclidean ball of radius~$R$, 
it is shown that 
complete polynomial 
complexity can in fact be achieved if~$R\leq c\sqrt{d}$. On the other hand, an exponential 
number of point evaluations is shown to be generally necessary for any algorithm as soon as~$R\geq C\sqrt{d}$ for constants~$C>c>0$. 
Importance sampling with a tail-matching proposal achieves the former, owing to a blessing of dimensionality. 
It is also shown that if strong concavity outside a ball is replaced by a distant dissipativity condition, then sampling guarantees must generally scale exponentially with~$d$ in essentially all parameter regimes. 
\end{abstract}

\section{Introduction}
In the literature, it is well-documented that basic 
Markov Chain Monte Carlo (MCMC) methods (e.g.~\cite{durmus2023asymptotic,MR2858447,MR1440273}) can be inefficient for sampling nonlogconcave target distributions~\cite{MR4417577,mangoubi2018d,MR3262493}. 
This shortcoming is observed both in practice and in theoretical guarantees, which are by now well-developed in both logconcave~\cite{pmlr-v83-c,gouraud2023hmc,MR4748799} and more general~\cite{MR4763254,MR4497240,MR4133372,chak2024r,pmlr-v178-c,pmlr-v195-f,MR4025861,MR4132634,MR4704569,schuh2024c} cases. 
To address this bottleneck, a large number of MCMC approaches specifically designed to address nonlogconcavity, especially metastability~\cite[Section~2.3.2.2]{MR2681239}, have been introduced, see~\cite{henin23,MR4630952,MR4152629,syed2024} and references within. 
From the theoretical viewpoint, a variety of analytical approaches have accompanied these contributions, but the assumptions are typically incomparable to those for basic MCMC. 
For example, in the line of work~\cite{surjanovic2024u,MR4412989,syed2024} on parallel tempering, convergence analysis is provided under an \textit{efficient local exploration} assumption. Under this condition, an advantage over other tempering methods is established, 
with theoretical improvement over local MCMC in high dimensions left open. 
As another example, in~\cite{MR4700257,MR2521882,MR2495560}, conditions on the target distribution are given to determine the scaling in dimension for the convergence of tempering and sequential Monte Carlo methods. These conditions depend on quantities that are based on a suitable partitioning of the state space into local modes, which are useful for analysis, but in general not expected to be known a priori or  well-scaled with respect to dimension. Likewise, in~\cite{MR3211000,guo2024p,NEURIPS2018_c6ede20e}, good estimates scaling polynomially in problem parameters are obtained, but under uniform structure, in the sense that the target is a mixture of uniform logconcave components, or that the target is of product form.

By contrast, guarantees for basic MCMC methods in both unimodal and multimodal cases, as referenced above, are thought to depend explicitly and polynomially on dimension under weaker settings. 
Here, a variety of quantitative assumptions have been used to specify multimodality. 
A basic general setting 
is where the target logdensity is strongly concave outside a ball around the origin with bounded Hessian. 
More precisely, these papers cover
target probability measures~$\mu$ on~$\mathbb{R}^d$ for which there exist~$L,m>0$,~$R\geq0$ and~$U\in C^2(\mathbb{R}^d)$ that satisfy~$\mu(dx) = (\int_{\mathbb{R}^d} e^{-U})^{-1} e^{-U(x)}dx$,~$L\geq m$ and
\begin{subequations}\label{set}
\begin{align}
\nabla U(0)&=0,\label{set0}\\
u^{\top} D^2 U (x) u &\geq m \abs{u}^2 & &\forall x\in\{y\in\mathbb{R}^d:\abs{y} > R\}, u\in\mathbb{R}^d,\label{set1}\\
\abs{D^2 U(x) u} &\leq L \abs{u} & &\forall x,u\in\mathbb{R}^d,\label{set2}
\end{align}
\end{subequations}
where~$D^2 U$ denotes the Hessian of~$U$. 
These assumptions represent a stylized but intuitive analytical setting. They are basic in the sense that~\eqref{set} is typically covered as a particular case (or as the main setting). 
For assumptions based on functional inequalities (e.g.~\cite{MR4763254,camrud2023second,pmlr-v178-c,pmlr-v195-f,MR4278799}), this setting is accounted for by a Holley-Stroock perturbation argument~\cite[Section~B.1]{MR4025861} or by Lyapunov functions~\cite{MR2386063}.
For works based on coupling analysis (e.g.~\cite{MR4133372,10.1214/23-EJP970,MR3933205,MR4132634,schuh2024c}), the main assumptions therein are similar, but in most cases strictly weaker, see any of the discussions following~\cite[Assumption~3]{MR4704569},~\cite[H2 (supplemental)]{NEURIPS2022_21c86d5b},~\cite[Assumption~1]{chak2024r}. 
The guarantees derived as a result are completely explicit in~$d,L,m,R$. In particular, they call for a number of point evaluations (or complexity) which is polynomially dependent on~$d$, but the dependence on~$LR^2$ is exponentially poor in general. 
This polynomial dependence on~$d$ is celebrated in~\cite{MR4025861}, championing the scalability of MCMC alongside the aforementioned
works. 
On the other hand, the exponential dependence on~$LR^2$ encapsulates at the theoretical level metastability of local MCMC methods. 

For known MCMC approaches specifically designed to overcome metastability, 
obtaining explicit sampling guarantees in the theoretical setting~\eqref{set} (that do not scale exponentially with either~$d$ or~$LR^2$, see in particular~\cite{vacher2025p}) remains open. 
In this direction, a result appeared in~\cite{NEURIPS2018_c6ede20e} which indicated that complete polynomial dependence jointly in~$d,L,m^{-1},R$ is impossible (see however Remark~\ref{remt}\ref{nipt} below). 

\subsection{Summary of results}\label{contri}
In this paper, 
it is shown that, under~\eqref{set}, 
polynomial sampling complexity jointly in~$d,L,m^{-1},R$ is possible to achieve 
if, for concreteness, the condition~$7\sqrt{L}R(1+L/m)^{3/2}\leq \sqrt{d-1}$ holds. This general setting includes the case where the inequality is saturated, so it otherwise calls for an exponential-in-$d$ number of point evaluations using previous theory. 
The polynomial complexity is attained by targeting a suitable modified distribution (see~\eqref{pidef} below) and correcting with importance sampling (IS). 
Moreover, it is shown that exponential complexity must in general be necessary in the regime~$\sqrt{L}R\geq 208\sqrt{d\ln(1+L/m)}$. 
Thus the condition on~$\sqrt{L}R$ for polynomial complexity is not improvable in terms of the order in~$d$ using any algorithm. 

It is also shown that if~\eqref{set1} is just slightly relaxed to only a distant dissipativity condition (see~\eqref{disseq} below), then the necessity of exponential complexity applies to virtually all regimes of the parameters in the condition, leaving little room for general polynomial guarantees outside of~\eqref{set} for any sampling method. 
Notably, 
this result continues to hold even with the additional assumption that the target is unimodal. 
The rest of this section is an elaboration of these contributions. 

The main 
tool in our analysis is a flattened, or truncated, probability distribution~$\pi$ of the original target~$\mu(dx)\propto e^{-U(x)}dx $ with the form
\begin{equation}\label{pidef}
\pi(dx) =(\textstyle \int_{\mathbb{R}^d}e^{-T\circ U})^{-1} e^{-T(U(x))}dx,
\end{equation}
where~$T:[\min_{\mathbb{R}^d}U,\infty)\rightarrow[\min_{\mathbb{R}^d}U,\infty)$ is a nondecreasing function satisfying
\begin{equation}\label{Tdef}
T(y) = \begin{cases}
y &\textrm{if } y \geq M+2c,\\
M+c &\textrm{if } y \leq M
\end{cases}
\end{equation}
for some~$c\in[0,\infty)$,~$M\in[\min_{\mathbb{R}^d}U-2c,\infty)$. 
The function~$T\circ U$ is locally flat 
and the same as~$U$ in the tails, thus \textit{tail-matching}. 
The value~$M$ is chosen to determine the flat region on which local modes from~$U$ are made to collapse. If~$M$ is too large, then clearly there is little value in considering~$\pi$, but it turns out that if~$M$ is only suitably large, 
then a self-normalized IS approach (see e.g.~\cite[Section~9]{mcbook} or~\cite{chak2023o}) 
on MCMC algorithms targeting~$\pi$ 
leads to polynomial complexity guarantees. 
A good choice of~$M$ is made explicit in Corollary~\ref{coco2} below (see also Remark~\ref{remo}\ref{rem21i} for reasoning) under a general Gaussian tail Assumption~\ref{A1}, and in~\eqref{Mdefint} below for the specific setting~\eqref{set}.
Note if~$c\neq 0$ and~$T,U$ are explicitly differentiable, then~$T\circ U$ is also explicitly so, thus MCMC methods applicable to~$\mu$ are also applicable to~$\pi$. 
The dependence on~$c$ of the bounds appearing below will be exponential, therefore~$c$ is thought throughout to be~$O(1)$ as~$d$ or any other parameters increase to infinity. 
A specific choice of~$T$ is given in Section~\ref{impl} and 
the extent to which~$T\circ U$ 
inherits smoothness properties of~$U$ 
is described in Proposition~\ref{cob2}. 

For the particular setting~\eqref{set}, the resulting tail-matching IS procedure to estimate~$\mu(\phi)=\int_{\mathbb{R}^d}\phi d\mu$ for any bounded measurable~$\phi:\mathbb{R}^d\rightarrow\mathbb{R}$ may be summarized as follows. 
\begin{enumerate}
\item\label{s1} In~\eqref{Tdef}, fix
\begin{equation}\label{Mdefint}
M = U(0) + LR^2/2.
\end{equation}
\item\label{s2} For some~$N\in\mathbb{N}\setminus\{0\}$, obtain~$N$ approximate samples~$(x_i)_{i\in[1,N]\cap\mathbb{N}}$ of~$\pi$ given by~\eqref{pidef} with an MCMC algorithm of choice.
\item\label{s3} Estimate~$\mu(\phi)$ with
\begin{equation}\label{introest}
\bar{\mu}(\phi):=\frac{\frac{1}{N}\sum_{i=1}^N\phi(x_i)e^{T(U(x_i))-U(x_i)}}{\frac{1}{N}\sum_{i=1}^Ne^{T(U(x_i))-U(x_i)}}.
\end{equation}
\end{enumerate}
Under the condition~$7\sqrt{L}R(1+L/m)^{3/2}\leq \sqrt{d-1}$, for any~$\epsilon\in (0,1]$, 
the number of point evaluations of~$(U,\nabla U)$ necessary in step~\ref{s2} (in this case with basic Langevin-type algorithms) to obtain~$\mathbb{E}[(\bar{\mu}(\phi) - \mu(\phi))^2]\leq \epsilon\max_x\abs{\phi(x)}^2$ depends polynomially on~$d,L,m,R,\epsilon$. 
This is the effort of Corollary~\ref{sob}, 
alongside the results of Section~\ref{impl} leading to Remark~\ref{jjq}. 
Namely, Corollary~\ref{sob} shows 
that~$\pi$ is close to~$\mu$ in the~$\chi^2$ sense (though note the logdensity values may differ significantly), 
whereas the results of Section~\ref{impl} show that~$\pi$ is (non-strongly) logconcave, thus accurately sampling from~$\pi$ is possible in polynomial time. 
Concrete Gaussian mixture examples satisfying~\eqref{set} are given in Section~\ref{gauex}. 
The condition~$7\sqrt{L}R(1+L/m)^{3/2}\leq \sqrt{d-1}$ 
is optimal up to a polynomial factor in~$(1+L/m)$ as mentioned (see again Remark~\ref{remo}\ref{rem3}). This optimality is the main goal in Section~\ref{bober}, in particular Theorem~\ref{comp}, where a counterexample is given in the regime where the radius~$R$ is too large. The quantitative intractability regime~$\sqrt{L}R\geq 208\sqrt{d\ln(1+L/m)}$ is derived in Remark~\ref{remt}\ref{rem2}. 

More generally, 
the alternative target~\eqref{pidef} is applicable 
under the Gaussian tail Assumption~\ref{A1} below on the values of~$U$, rather than~$D^2U$ as in~\eqref{set}. 
In this case, 
only the choice of~$M$ in step~\ref{s1} changes, which is instead based on the constants with which Assumption~\ref{A1} is satisfied. 
Under Assumption~\ref{A1} and a quantitative condition~\eqref{cocoa} on the constants therein, Corollary~\ref{coco2} shows that, for suitably chosen~$M$,~$\pi$ is a suitable distribution to sample from instead of~$\mu$, in the sense that 
IS with proposal~$\pi$ is in principle well-justified. 
Assumption~\ref{A1} and condition~\eqref{cocoa} are verified in the context of large-scale Bayesian neural networks in Section~\ref{bnn}. 
Note that sampling from~$\pi$ (step~\ref{s2} above) may be carried out with algorithms of choice, 
for example parallel tempering or sequential Monte Carlo samplers, or more recent methodologies~(e.g.~\cite{MR4417577,noble2024l,vacher2025p}) based on reverse diffusions or normalizing flows. 
Intuitively, the alternative target~\eqref{pidef} 
explicitly informs MCMC algorithms of regions of the state space that can be safely flattened, then corrected with importance sampling, thereby preventing unnecessary exploration of these regions (if any). 
See also the paragraph preceding Section~\ref{bober} for a potential use beyond basic IS. 
However, no empirical evidence is given for its practical benefits in this paper and it is considered primarily for its analytical properties here. 
In particular, this applicability of~\eqref{pidef} 
leads to the question as to the availability of scalable guarantees outside of~\eqref{set}. In this Gaussian tail setting, forfeiting~\eqref{set1} means that~$\pi$ is in many cases not logconcave, which means satisfactory guarantees for sampling from~$\pi$ are not available, in contrast to the situation under~\eqref{set}. 

In the direction of guarantees for sampling~$\mu$ under such weak conditions, it is shown in Section~\ref{cocer} that if~\eqref{set1} is replaced just by distant dissipativity\footnote{The term `distant dissipativity' is used slightly differently in some works (e.g.~\cite{kern}).}, as in
\begin{equation}\label{disseq}
\nabla U (x)\cdot x \geq \alpha \abs{x}^2 - \beta \qquad \forall x\in\mathbb{R}^d\setminus B_{\!\!\sqrt{\beta/\alpha\,}},
\end{equation}
for some~$\alpha>0$,~$\beta\geq 0$, then it is impossible to conclude that~$\mu$ 
can be sampled in polynomial time in essentially any regime of the parameters~$\alpha,\beta$. 
Distant dissipativity is considered (under more general conditions or otherwise) in for example~\cite{MR4749765,MR4313846,pmlr-v134-e,subopt,MR4531091,pmlr-v65-r,pmlr-v75-t,MR3555050,MR4534450,NEURfeb}, with analysis resulting in inexplicit or unscalable rates. 
Note that~\eqref{disseq} is weaker than~\eqref{set1} in the sense of (the proof of) Proposition~\ref{cob}, but stronger than Assumption~\ref{A1} in the sense of Proposition~\ref{diss}. 
The main result in Section~\ref{cocer} 
is Theorem~\ref{comp2}, 
which concerns the strongest possible value for~$\beta$ (that is~$\beta=0$, which implies unimodality in the sense that~$U$ admits a unique minimum) and~$\alpha$ proportional to~$L$ from~\eqref{set2}. 
As a secondary consequence of Theorem~\ref{comp2}, since in the case~$\beta=0$ Corollary~1.6 (1) in~\cite{MR2386063} may be applied, any Poincar\'e constant derived directly therein must always scale exponentially with~$d$, see Remark~\ref{cober2}.

The rest of the paper is organized as follows. Section~\ref{mse} gives the main results on the performance of IS estimators given i.i.d. samples from the proposal~\eqref{pidef}. In particular, the general explicit condition~\eqref{cocoa} on~$U$ is given in Corollary~\ref{coco2} under which tail-matching IS is valid. 
In Section~\ref{intr}, the opposing counterexamples 
are given for which sampling guarantees necessitate an exponential number of point evaluations. In Section~\ref{impl}, it is shown that, given convexity outside a ball and the suitable quantitative condition, some implementable IS algorithms achieve polynomial complexity. 
Section~\ref{examples} 
presents concrete examples where the results in Sections~\ref{mse} and~\ref{impl} may be applied.
\paragraph*{Notation}
The notation~$\mathbb{N}$ for natural numbers includes~$0$. 
The closed Euclidean ball of radius~$r\geq 0$ around~$x\in\mathbb{R}^d$ is denoted by~$B_r(x)$, with~$B_r:=B_r(0)$. The notation~$\lambda$ is used for the Lebesgue measure on~$\mathbb{R}^d$. The projection of a point~$x\in\mathbb{R}^d$ to the set~$\bar{B}\subset\mathbb{R}^d$ is denoted~$P_{\bar{B}}x := \argmin_{y\in \bar{B}}\abs{x-y}$ when it is unique and if~$\bar{B}=B_r$, then~$P_rx := P_{B_r}x$. The gamma function~$\Gamma$ is given by~$\Gamma(x) = \int_0^{\infty}t^{x-1}e^{-t}dt$ for all~$x>0$. The indicator function for a set~$S$ is denoted~$\mathds{1}_S$. The~$d$-by-$d$ identity matrix is denoted~$I_d$. 
A function is called smooth if it is infinitely differentiable (not to be confused with~$L$-smoothness). 
The standard mollifier on~$\mathbb{R}$ is denoted~$\varphi:\mathbb{R}\rightarrow[0,\infty)$, that is,~$\varphi(x)=(\int_{-1}^1 \exp(-1/(1-\abs{y}^2))dy)^{-1}\exp(-1/(1-\abs{x}^2))$ for~$x\in(-1,1)$ and~$\varphi(x)=0$ elsewhere. For~$m_1,m_2\in\mathbb{N}\setminus\{0\}$ and matrices~$A,B\in\mathbb{R}^{m_1\times m_2}$, the operator norm of~$A$ is denoted~$\abs{A} = \sup_{u\in\mathbb{R}^{m_2},\abs{u}=1}\abs{Au}$ 
and the double dot product is denoted~$A:B=\sum_{ij}A_{ij}B_{ij}$. 
Throughout, we denote~$Z=\int_{\mathbb{R}^d} e^{-U}$. 

\section{Chi-square divergence estimate}\label{mse}

The starting point for our analysis of IS estimators is the main result in~\cite[Theorem~2.1]{MR3696003}. This result is restated here 
for~$\pi$. 
\begin{theorem}
\label{stu}
Let~$c\in(0,\infty)$,~$M\in[\min_{\mathbb{R}^d}U,\infty)$ 
and let~$\pi$ be a probability distribution with density given by~\eqref{pidef} 
for some~$C^2$ nondecreasing~$T:[\min_{\mathbb{R}^d}U,\infty)\rightarrow[\min_{\mathbb{R}^d}U,\infty)$ 
satisfying~\eqref{Tdef} and~$T''\geq 0$. 
Let~$(u^n)_{n\in\mathbb{N}}$ be an i.i.d. sequence of~$\mathbb{R}^d$-valued r.v.'s with~$u^n\sim \pi$. 
For any~$N\in\mathbb{N}\setminus\{0\}$, it holds that
\begin{equation*}
\sup_{\phi}|\mathbb{E}[\mu^N(\phi) - \mu(\phi)]|\leq \frac{12}{N}\rho,\qquad
\sup_{\phi}\mathbb{E}[(\mu^N(\phi) - \mu(\phi))^2]\leq \frac{4}{N}\rho,
\end{equation*}
where the suprema are over all measurable~$\phi:\mathbb{R}^d\rightarrow\mathbb{R}$ with~$\sup_x\abs{\phi(x)} \leq 1$,~$\mu^N(\phi)$ denotes~$\mu^N(\phi) = (\sum_{n=1}^N \phi(u^n) e^{T\circ U(u^n)-U(u^n)})(\sum_{m=1}^N e^{T\circ U(u^m)-U(u^m)})^{-1}$ and
\begin{equation}\label{rhob}
\rho = \pi(e^{2T\circ U-2U})/(\pi(e^{T\circ U-U}))^2 .
\end{equation}
\end{theorem}
The main effort in this section is to control~$\rho$. Note that~$\rho-1$ is the~$\chi^2$ divergence between the target~$\mu$ and proposal~$\pi$, see e.g.~\cite[equation~(2.4)]{MR3696003}. 
The results in Section~\ref{msemain} below show that, under a general Gaussian tail Assumption~\ref{A1} and an explicit condition on the constants therein, the self-normalized IS estimator based on exact samples from the flattened proposal~\eqref{pidef} is well-behaved, in the sense that~$N$ in Theorem~\ref{stu} need not scale with any problem parameters to achieve good accuracy. 
The following Lemma~\ref{base} gives the basic argument by which this control on~$\rho$ is obtained.
\begin{lemma}\label{base}
Assume the setting and notations of Theorem~\ref{stu}. 
It holds that
\begin{align}
\rho &\leq \frac{1}{Z^2}\bigg[\lambda(U\leq M)^2e^{-2\min_{\mathbb{R}^d} U}  + e^c\bigg(\int_{U > M}e^{-U}\bigg)^{\!\!2} \nonumber \\
&\quad+ \lambda(U\leq M)(e^{-2M}+e^{c- 2\min_{\mathbb{R}^d}U}) \int_{U > M} e^{M-U}\bigg].\label{rhob2}
\end{align}
\end{lemma}
The interest in Lemma~\ref{base} is that for general classes of nonconvex functions~$U$ and well-chosen~$M$, the right-hand side of~\eqref{rhob2} can be controlled by estimates on~$\lambda(U\leq M)$ and the radial growth of~$U$ at infinity. 
In addition, the squared integral on the right-hand side of~\eqref{rhob2} always appears in a corresponding expression for~$Z^2$, so that if this squared integral is the dominating term in the square brackets, then~$\rho$ is well approximated by~$1$.
\begin{proof}[Proof of Lemma~\ref{base}]
Let~$Z_{\pi}$ be the normalization constant for~$\pi$, that is,~$Z_{\pi} = \int_{\mathbb{R}^d} e^{-T\circ U}$. By definition of~$T$,
it holds that
\begin{align*}
&\pi(e^{2T\circ U-2U})\cdot Z_{\pi}^2 \\
&\quad= \bigg(\int_{\mathbb{R}^d} e^{T\circ U-2U} \bigg)\bigg(\int_{\mathbb{R}^d} e^{-T\circ U}\bigg)\\
&\quad= \bigg(\int_{U\leq M}  e^{T\circ U-2U} + \int_{U> M}  e^{T\circ U-2U}\bigg) \cdot \bigg( \int_{U\leq M}  e^{-T\circ U} + \int_{U> M}  e^{-T\circ U}\bigg)\\
&\quad\leq \bigg( \lambda(U\leq M)  e^{M+c-2\min_{\mathbb{R}^d}\!U} + \int_{U> M}e^{c-U}\bigg) \\
&\qquad\cdot \bigg(\lambda (U\leq M) e^{-M-c} + \int_{U> M} e^{-U}\bigg),
\end{align*} 
from which the conclusion follows.
\end{proof}

\subsection{Main estimates}\label{msemain}

The assumption required to establish the results in this section is summarized in the following Assumption~\ref{A1}. 
\begin{assumption}\label{A1}
There exist~$c_U,\mathcal{R}\geq 0$,~$L\geq m>0$ and~$x^*\in B_{\mathcal{R}}$ such that
it holds 
for any~$x\in\mathbb{R}^d$ that
\begin{equation}\label{A1eq}
\frac{m}{2}(\abs{x}-\mathcal{R} )^2\mathds{1}_{\mathbb{R}^d\setminus B_{\mathcal{R}}}(x) \leq U(x) - \min_{\mathbb{R}^d}U \leq  c_U + \frac{L}{2}|x-x^*|^2.
\end{equation}
\end{assumption}
Assumption~\ref{A1} is satisfied with~$c_U=0$ when~$U$ is strongly convex outside some Euclidean ball and~$U$ is differentiable with a gradient that is globally Lipschitz. This is made precise in Proposition~\ref{cob}. 
More generally, given only a distant dissipativity condition, it is verified in Proposition~\ref{diss} that the left-hand inequality of~\eqref{A1eq} is satisfied, so that all of Assumption~\ref{A1} is satisfied for example if in addition~$U$ admits a globally Lipschitz gradient. From a practical viewpoint, Assumption~\ref{A1} may easily be satisfied for Bayesian inference problems with Gaussian priors, see Section~\ref{bnn}. 

Assumption~\ref{A1} offers a basic setting where Lemma~\ref{base} may be applied to obtain an explicit (in particular in~$d$) and favourable estimate on~$\rho$. 
Specifically, the estimate is bounded above by an absolute constant as~$d\rightarrow\infty$ (marginally in~$d$) and polynomial in~$\mathcal{R}$ as~$\mathcal{R}\rightarrow\infty$ (marginally in~$\mathcal{R}$). 

In the next Theorem~\ref{coco}, this explicit estimate on~$\rho$ is given under Assumption~\ref{A1}. For small enough~$\mathcal{R}$ (in relation to~$m,L,d,M$), this estimate implies a constant (in particular in~$d$) bound on~$\rho$, which leads to the announced results. This consequence is stated precisely in Corollary~\ref{coco2}. In this latter result, the threshold~$M$ from~\eqref{Tdef} is fixed w.r.t the other parameters in a way such that the importance sampler associated to~$M$ is effective. In Theorem~\ref{coco}, a slightly sharper bound is also given in the stronger case~\eqref{set}.

Theorem~\ref{coco} and the subsequent results are vacuous when~$d=1$, but the statements and proofs can be modified to accommodate this case. The focus is on the high-dimensional case, so the~$d=1$ case is omitted. 
\begin{theorem}\label{coco}
Assume the setting and notations of Theorem~\ref{stu} and~$d\geq 2$. 
If Assumption~\ref{A1} holds, then it holds that
\begin{equation}\label{frb}
\rho \leq (2e^c) \vee\bigg(\frac{e^{2c_U}(2+e^c)(L\bar{R}^2)^d}{2^{d-1}(\Gamma(d/2+1))^2} + \frac{e^{2c_U}(1+e^c)2^d}{e} \sum_{i=0}^{d-1}\bigg(\frac{e\cdot L\bar{R}^2(L/m)}{d-1}\bigg)^{d-\frac{1}{2}-\frac{i}{2}}\bigg),
\end{equation}
where
\begin{equation}\label{Rbdef}
\bar{R} = \mathcal{R} + \sqrt{\frac{2}{m}\bigg(M-\min_{\mathbb{R}^d} U\bigg)}.
\end{equation}
Moreover, if instead of Assumption~\ref{A1}, there exist~$L\geq m>0$,~$R\geq0$ such that~\eqref{set} hold and~$M\geq U(0) - LR^2/2$, 
then the same inequality~\eqref{frb} holds with~$c_U=0$ and
\begin{equation}\label{Rbdef2}
\bar{R} = R \bigg(1+\frac{L}{m}\bigg) + \sqrt{\frac{L^2R^2}{m^2} - \frac{2}{m}\bigg(U(0)-\frac{LR^2}{2}\bigg) + \frac{2M}{m}}.
\end{equation}
\end{theorem}
\begin{proof}
The various terms appearing on the right-hand side of~\eqref{rhob2} are estimated in terms of the quantities in Assumption~\ref{A1} and~\eqref{set} respectively. 
In case Assumption~\ref{A1} and~\eqref{Rbdef} hold, the left-hand bound in~\eqref{A1eq} implies
\begin{equation*}
U(x)-\min_{\mathbb{R}^d}U\geq (m/2)(\abs{x}-\bar{R}+\bar{R}-\mathcal{R})^2\geq (m/2)((\abs{x}-\bar{R})^2+(\bar{R}-\mathcal{R})^2)\quad\forall x\in\mathbb{R}^d\setminus B_{\bar{R}}
\end{equation*}
and so 
\begin{equation}\label{uma}
U(x) 
\geq  M+\frac{m}{2}(\abs{x}-\bar{R})^2\qquad\forall x\in \mathbb{R}^d\setminus B_{\bar{R}}.
\end{equation}
In case~\eqref{set} and~\eqref{Rbdef2} hold instead of Assumption~\ref{A1} and~\eqref{Rbdef}, by the fundamental theorem of calculus, it holds for any~$x\in\mathbb{R}^d\setminus B_{\bar{R}}$ that
\begin{align*}
U(x) &\geq U(Rx/\abs{x}) + (x-Rx/\abs{x})\cdot \nabla U(Rx/\abs{x}) + m\abs{x-Rx/\abs{x}}^2/2\\
&\geq U(0) - LR^2/2 -LR(\abs{x}-R) + m(\abs{x}-R)^2/2.
\end{align*}
For~$\abs{x}=\bar{R}$, the right-hand side is equal to~$M$ (the definition~\eqref{Rbdef2} of~$\bar{R}$ is made for this reason). Moreover, it is easy to see by taking derivatives w.r.t.~$\abs{x}$ that, for~$\abs{x}\geq \bar{R}$, the right-hand side is increasing in~$\abs{x}$ with second derivative at least~$m$. Therefore~\eqref{uma} also holds in this case.

In both cases,~\eqref{uma} implies
\begin{equation}\label{lma}
\lambda(U\leq M) \leq \lambda(B_{\bar{R}}) = \pi^{\frac{d}{2}}\bar{R}^d/\Gamma(d/2+1).
\end{equation}
For the integral appearing in~\eqref{rhob2}, it holds that
\begin{equation}\label{vma}
\int_{U > M} e^{-U} = \int_{B_{\bar{R}}\cap \{U>M\}}e^{-U} + \int_{\mathbb{R}^d\setminus B_{\bar{R}}} e^{-U}, 
\end{equation}
where the first term on the right-hand side of~\eqref{vma} satisfies
\begin{equation}\label{vna}
\int_{B_{\bar{R}}\cap \{U>M\}}e^{-U} \leq  \frac{\lambda(B_{\bar{R}})}{ e^M} = \frac{\pi^{\frac{d}{2}}\bar{R}^d }{\Gamma(d/2+1)e^M}
\end{equation}
and by~\eqref{uma} 
the second term on the right-hand side of~\eqref{vma} satisfies 
\begin{equation}\label{umb}
\int_{\mathbb{R}^d\setminus B_{\bar{R}}} e^{-U}\leq  e^{-M}\int_{\mathbb{R}^d\setminus B_{\bar{R}}} e^{-\frac{m}{2}(\abs{x}-\bar{R})^2} dx.
\end{equation}
By a spherical coordinate transform, the integral on the right-hand side of~\eqref{umb} satisfies
\begin{align}
\int_{\mathbb{R}^d\setminus B_{\bar{R}}} e^{-\frac{m}{2}(\abs{x}-\bar{R})^2} dx &= \frac{2\pi^{\frac{d}{2}}}{\Gamma(d/2)}\int_{\bar{R}}^{\infty} r^{d-1}e^{-\frac{m}{2}(r-\bar{R})^2} dr\nonumber\\
&=\frac{2\pi^{\frac{d}{2}}}{\Gamma(d/2)} \int_0^{\infty} (r+\bar{R})^{d-1} e^{-\frac{m}{2}r^2} dr\nonumber\\
&=\frac{2\pi^{\frac{d}{2}}}{\Gamma(d/2)} \sum_{i=0}^{d-1}\begin{pmatrix} d-1\\ i \end{pmatrix} \bar{R}^{d-1-i} \int_0^{\infty} r^i e^{-\frac{m}{2}r^2} dr\nonumber\\
&\leq \frac{\sqrt{2}\pi^{\frac{d+1}{2}}}{\Gamma(d/2)}  \sum_{i=0}^{d-1}\begin{pmatrix} d-1\\ i \end{pmatrix} \bar{R}^{d-1-i} m^{-\frac{i+1}{2}} (i-1)!!,\label{vmb}
\end{align}
where~$(-1)!!:=1$. 
The sum on the right-hand side of~\eqref{vmb} can be rewritten as
\begin{equation}\label{bmb}
S:=\sum_{i=0}^{d-1}\begin{pmatrix} d-1\\ i \end{pmatrix} \bar{R}^{d-1-i} m^{-\frac{i+1}{2}} (i-1)!! = (d-1)!\sum_{i=0}^{d-1}\frac{\bar{R}^{d-1-i} m^{-\frac{i+1}{2}} }{(d-1-i)!i!!}. 
\end{equation}
Note by~$n! \in [(2\pi)^{1/2}n^{1/2}(n/e)^n, en^{1/2}(n/e)^n]$, it holds for any even~$i\in\mathbb{N}\setminus\{0\}$ that
\begin{equation*}
i!! = 2^{i/2}(i/2)! \geq 2^{i/2} \cdot (2\pi)^{1/2}e^{-i/2}(i/2)^{(i+1)/2} \geq  \pi^{1/2}e^{-i/2}i^{(i+1)/2}
\end{equation*}
and for any odd~$i\in\mathbb{N}$ that
\begin{align*}
i!! &= \frac{(i+1)!}{2^{(i+1)/2}((i+1)/2)!} \\
&\geq \frac{(2\pi)^{1/2}(i+1)^{1/2}(i+1)^{i+1}e^{-i-1}}{2^{(i+1)/2} e^{1-(i+1)/2} ((i+1)/2)^{1/2} ((i+1)/2)^{(i+1)/2}}\\
&= \frac{2\pi^{1/2}}{e}\bigg(\frac{i+1}{e}\bigg)^{\frac{i+1}{2}},
\end{align*}
which together imply
\begin{equation}\label{dfb}
i!!\geq 2\pi^{1/2}e^{-(i+1)/2}i^{i/2}\qquad\forall i\in\mathbb{N}\setminus\{0,1\},
\end{equation}
where we have used~$(i+1)^{(i+1)/2}= ((1+1/i)^i\cdot i^i)^{1/2}\cdot (i+1)^{1/2} \geq 2^{1/2}i^{i/2}\cdot 2$ for odd~$i\geq 3$. 
Therefore, the denominator on the right-hand side of~\eqref{bmb} satisfies for any~$i\in[2,d-1)\cap\mathbb{N}$ that
\begin{align}
(d-1-i)!i!! &\geq (2\pi)^{1/2}(d-1-i)^{1/2}((d-1-i)/e)^{d-1-i}\cdot 2\pi^{1/2}i^{i/2}e^{-(i+3)/2}\nonumber \\
&= \sqrt{8}\,\pi e^{(i-1)/2-d} (d-1-i)^{d-1/2-i}\cdot i^{i/2}\nonumber\\
& \geq e\cdot e^{(1+i)/2-d} (d-1-i)^{d-1-i}\cdot i^{i/2}.\label{jdw}
\end{align}
For any~$i\in\{0,1\}$, if~$d\geq 3$, then the left-hand side is bounded below by the right-hand side by~$n!\geq (2\pi)^{1/2}n^{1/2}(n/e)^n$. For~$d=2$, the same holds by inspection for~$i\in\{0,1\}$. For~$i=d-1$ (when~$d\geq 3$), the same again holds by~\eqref{dfb}.
Denote~$\bar{d}=d-1$. To estimate the right-hand side of~\eqref{jdw}, note for any~$\bar{\lambda}\in[0,1]$, it holds that
\begin{equation*}
(\bar{d}(1-\bar{\lambda}))^{\bar{d}(1-\bar{\lambda})}\cdot(\bar{d}\bar{\lambda})^{\bar{d}\bar{\lambda}/2} = \big((1-\bar{\lambda})^{1-\bar{\lambda}}\cdot\bar{\lambda}^{\bar{\lambda}/2}\big)^{\bar{d}}\cdot \bar{d}^{\bar{d}(1-\bar{\lambda}/2)}\geq 2^{-\bar{d}} \cdot \bar{d}^{\bar{d}(1-\bar{\lambda}/2)}.
\end{equation*}
Applying this on the right-hand side of~\eqref{jdw} then inserting into~\eqref{bmb} 
yields
\begin{equation}\label{Sb}
S\leq (2e)^{d-1}(d-1)!\sum_{i=0}^{d-1}\frac{\bar{R}^{d-1-i}m^{-\frac{i+1}{2}}}{e^{\frac{i+1}{2}}(d-1)^{d-1-\frac{i}{2}}} .
\end{equation}
By substituting into~\eqref{vmb}, it holds that
\begin{equation}\label{jqp}
\int_{\mathbb{R}^d\setminus B_{\bar{R}}} e^{-\frac{m}{2}(\abs{x}-\bar{R})^2} dx \leq \frac{\pi^{\frac{d+1}{2}}}{\Gamma(d/2)} \cdot \sqrt{2}(2e)^{d-1}(d-1)!\sum_{i=0}^{d-1}\frac{\bar{R}^{d-1-i}m^{-\frac{i+1}{2}}}{e^{\frac{i+1}{2}}(d-1)^{d-1-\frac{i}{2}}}
\end{equation}
In order to simplify the right-hand side of~\eqref{jqp}, note by Legendre's duplication (which gives~$\Gamma(d/2)\Gamma((d+1)/2) = 2^{1-d}\sqrt{\pi}\Gamma(d)$) and Gautschi's inequality (which gives~$\Gamma((d+1)/2)/\Gamma(d/2+1)< (2/d)^{1/2}$), the Gamma function satisfies
\begin{equation*}
\frac{(d-1)!}{\Gamma(d/2)\Gamma(d/2+1)} = \frac{\Gamma(d)}{\Gamma(d/2)\Gamma(d/2+1)}  = \frac{2^{d-1}\Gamma((d+1)/2)}{\sqrt{\pi}\Gamma(d/2+1)} < \frac{2^d}{\sqrt{2\pi d\,}}.
\end{equation*}
Substituting into~\eqref{jqp} after multiplying by~$1=\frac{\Gamma(d/2+1)}{(L\bar{R})^d}\cdot\frac{(L\bar{R})^d}{\Gamma(d/2+1)}$ yields (defining the following expression with the limit in case~$\bar{R}=0$) 
\begin{equation*}
\int_{\mathbb{R}^d\setminus B_{\bar{R}}} e^{-\frac{m}{2}(\abs{x}-\bar{R})^2} dx \leq\frac{\Gamma(d/2+1)}{(L\bar{R})^d} \cdot \frac{2(4e)^{d-1}\pi^{\frac{d}{2}}\sqrt{d-1}}{\sqrt{d}}\sum_{i=0}^{d-1} \frac{(\sqrt{L}\bar{R})^{2d-1-i}(L/m)^{\frac{i+1}{2}}}{e^{\frac{i+1}{2}}(d-1)^{d-\frac{1}{2}-\frac{i}{2}}}.
\end{equation*}
Together with~\eqref{umb},~\eqref{vna},~\eqref{vma},~\eqref{lma}, and~$(L/m)^{(i+1)/2}\leq (L/m)^{d/2}\leq (\sqrt{L/m})^{2d-1-i}$ for all~$i\in[0,d-1]\cap\mathbb{N}$, this implies
\begin{align}
&\lambda(U\leq M)\int_{U>M}e^{M-U} \nonumber \\
&\quad\leq \frac{\pi^d\bar{R}^{2d}}{(\Gamma(d/2+1))^2} + \frac{2(4e)^{d-1}\pi^d}{L^d} \sum_{i=0}^{d-1} \frac{(\sqrt{L}\bar{R}\sqrt{L/m})^{2d-1-i}}{e^{\frac{i+1}{2}}(d-1)^{d-\frac{1}{2}-\frac{i}{2}}}.\label{jue}
\end{align}
Gathering~\eqref{lma},~\eqref{jue},~$e^{\min_{\mathbb{R}^d}U - M}\leq 1$ and substituting into~\eqref{rhob2} yields
\begin{equation}\label{kkd}
\rho \leq \frac{1}{(Ze^{\min_{\mathbb{R}^d}U})^2} \bigg( \rho_0 + e^c\bigg(\int_{U>M}e^{-U+\min_{\mathbb{R}^d}U}\bigg)^{\!\!2}\,\bigg),
\end{equation}
where
\begin{equation*}
\rho_0:= (2+e^c)\frac{\pi^d\bar{R}^{2d}}{(\Gamma(d/2+1))^2} + (1+e^c)\frac{2(4e)^{d-1}\pi^d}{L^d} \sum_{i=0}^{d-1} \frac{(\sqrt{L}\bar{R}\sqrt{L/m})^{2d-1-i}}{e^{\frac{i+1}{2}}(d-1)^{d-\frac{1}{2}-\frac{i}{2}}}.
\end{equation*}
By definition, it holds that~$Ze^{\min_{\mathbb{R}^d} U}= \int_{U\leq M} e^{-U+\min_{\mathbb{R}^d} U} + \int_{U>M} e^{-U+\min_{\mathbb{R}^d} U}$. Two possibilities for the value of the latter integral are considered. In case it holds that
\begin{equation}\label{kd0}
e^c\bigg( \int_{U>M} e^{-U+\min_{\mathbb{R}^d}U}\bigg)^2 \geq \rho_0,
\end{equation}
inequality~\eqref{kkd} implies~$\rho\leq 2e^c$. In the other case where~\eqref{kd0} does not hold, note first that both the assumption~\eqref{A1eq} and~\eqref{set} (with~$c_U=0$) imply~$Z 
\geq 
e^{-c_U-\min_{\mathbb{R}^d}U}(2\pi/L)^{\frac{d}{2}}$. 
Therefore inequality~\eqref{kkd} and the negation of~\eqref{kd0} imply 
\begin{equation*}
\rho \leq e^{2c_U}(L/(2\pi))^d \cdot 2\rho_0,
\end{equation*}
which concludes the proof. 
\end{proof}
\begin{corollary}\label{coco2}
Let~$\hat{c}\geq 1$. 
Suppose~$d\geq 2$ and Assumption~\ref{A1} holds with~$c_U,m,L,\mathcal{R}$ satisfying
\begin{equation}\label{cocoa}
\frac{L}{m}\bigg(\sqrt{L}\mathcal{R} + \bigg(\frac{L}{m}(4c_U + 5L\mathcal{R}^2)\bigg)^{\!\!\frac{1}{2}} \,\bigg)^2  \leq \frac{d-1}{5e\hat{c}}.
\end{equation}
Assume the setting and notations of Theorem~\ref{stu}. 
Let~$M=U(0) + c_U + 2L\mathcal{R}^2$. 
The constant~$\bar{R}$ defined by~\eqref{Rbdef} satisfies~$L\bar{R}^2(L/m)\leq (d-1)/(5e\hat{c})$ and 
it holds that
\begin{equation}\label{cocob}
\rho \leq (2e^c) \vee \bigg(\frac{2e^{2c_U}(2+e^c)}{e(5\hat{c})^d\sqrt{d\pi/2\,}} + \frac{e^{2c_U}(1+e^c)}{(5\hat{c}/4)^{d/2}(1-\sqrt{4/(5\hat{c})}\,)e}\bigg).
\end{equation}
\end{corollary}
\begin{remark}\label{remo}
\begin{enumerate}[label=(\roman*)]
\item\label{rem21i} 
The value fixed for~$M$ in Corollary~\ref{coco2} is a worst case estimate for~$\max_{B_{\mathcal{R}}}U$ based on one evaluation of~$U$ at zero and given Assumption~\ref{A1}. 
At the intuitive level, this choice of~$M$ collapses all of the multimodality posed in~$B_{\mathcal{R}}$ and condition~\eqref{cocoa} enforces the negligibility of this multimodality for sampling~$\mu$. 
Corollary~\ref{coco2} proves this negligibility by asserting that the consequent definition~\eqref{Rbdef} for~$\bar{R}$ leads to a favourable estimate of~$\rho$. 
\item\label{crem} Recall from Assumption~\ref{A1} that~$c_U$ measures, loosely speaking, the variation of~$U$ exceeding~$L$-smoothness. Non-unit values for~$\hat{c}$ allow~$c_U$ to increase with~$d$ whilst keeping a scalable estimate on~$\rho$. 
For example, if~$c=1$ and~\eqref{cocoa} holds with~$\hat{c} = e^{1/(10e)}$, then~\eqref{cocoa} implies~$c_U\leq d/(20e)$, which, together with~\eqref{cocob}, implies for~$d\geq 2$ that~$\rho\leq 10$.
In the case where~$U$ admits a globally Lipschitz gradient, we may take~$c_U=0$ for simplicity, in which case taking~$\hat{c}=1$ suffices to obtain a satisfactory bound on~$\rho$. 
However, it can be more natural to consider~$c_U\neq0$ in any case, see Section~\ref{examples}. 
\item \label{rem3} 
In the~$L$-smooth case with~$c_U=0$ and~$\hat{c}=1$, inequality~\eqref{cocoa} is satisfied if
\begin{equation}\label{reeq}
\sqrt{L}\mathcal{R}(1+L/m)\leq (1/10)\sqrt{d-1}.
\end{equation}
The scaling required on the left-hand side of~\eqref{reeq} w.r.t.~$\kappa:=L/m$ seems unavoidable for the importance sampler presented here. In contrast to the scaling in~$d$, this will lead, together with Proposition~\ref{cob} below, to a scaling in~$\kappa$ that matches the intractability region in Theorem~\ref{comp} only up to a polynomial factor as announced, see Remark~\ref{remt}\ref{rem2}. 
Note that under~\eqref{set}, fixing~\eqref{Mdefint}, instead of the value for~$M$ inferred by Corollary~\ref{coco2} and Proposition~\ref{cob}, then tailoring the arguments leads to a slight improvement in scaling. This is the content of Corollary~\ref{sob}.
\end{enumerate}
\end{remark}
\begin{proof}[Proof of Corollary~\ref{coco2}]
From Theorem~\ref{coco}, it holds by Gautschi's inequality, Legendre's duplication and~$n!\geq e(n/e)^n$ that
\begin{equation}\label{jjv}
\rho \leq (2e^c) \vee\bigg(\frac{2e^{2c_U}(2+e^c)(eL\bar{R}^2)^d}{ed^d\sqrt{d\pi/2\,}} + \frac{e^{2c_U}(1+e^c)}{e} \sum_{i=0}^{d-1}\bigg(\frac{4e\cdot L\bar{R}^2(L/m)}{d-1}\bigg)^{d-\frac{1}{2}-\frac{i}{2}}\bigg),
\end{equation}
where~$\bar{R}$ is given by~\eqref{Rbdef}. 
In order to estimate~$\bar{R}$, 
note that the right-hand bound of~\eqref{A1eq} implies
\begin{equation*}
U(0)\leq \min_{\mathbb{R}^d} U + c_U + L\mathcal{R}^2/2.
\end{equation*}
Combining this with~\eqref{cocoa} and the assumption on~$M$ yields
\begin{equation*}
L\bar{R}^2(L/m) \leq (L^2/m)\big(\mathcal{R} + \big((2/m)(2c_U + 5L\mathcal{R}^2/2)\big)^{\frac{1}{2}}\big)^2 \leq (d-1)/(5e\hat{c}), 
\end{equation*}
which may be substituted into~\eqref{jjv} to obtain the assertion.
\end{proof}

\subsection{Application to known assumptions}

In what remains of this section, it is verified that the quantitative nonconvexity assumptions used in the literature as stated in the introduction are stronger than Assumption~\ref{A1}. 

\begin{prop}\label{diss}
Suppose~$U\in C^1$. If there exist~$\alpha>0$,~$\beta \geq 0$ such that~\eqref{disseq} holds, 
then the left-hand inequality in~\eqref{A1eq} is satisfied with~$m=\alpha$ and~$\mathcal{R}=\sqrt{\beta/\alpha}$. If in addition~$\nabla U$ is~$\bar{L}$-Lipschitz, then all of Assumption~\ref{A1} holds with also~$c_U=0$,~$L=\bar{L}$.
\end{prop}
\begin{proof}
Let~$m=\alpha$,~$\mathcal{R}=\sqrt{\beta/\alpha}$. For any~$x\in\mathbb{R}^d\setminus \{0\}$, denote~$\hat{x}=x/\abs{x}$. By the fundamental theorem of calculus, inequality~\eqref{disseq} implies for~$x\in\mathbb{R}^d\setminus B_{\mathcal{R}}$ that
\begin{align*}
U(x) &= U(P_{\mathcal{R}}x) + \int_0^1 \nabla U(\bar{\lambda}(x-P_{\mathcal{R}}x) + P_{\mathcal{R}}x) d\bar{\lambda} \cdot (x-P_{\mathcal{R}}x)\\
&= U(P_{\mathcal{R}}x) + \int_0^1 \nabla U\big(\hat{x}(\bar{\lambda}(\abs{x}-\mathcal{R}) + \mathcal{R})\big) \\
&\quad\cdot \hat{x}(\bar{\lambda}(\abs{x}-\mathcal{R})+\mathcal{R})(\abs{x}-\mathcal{R})/(\bar{\lambda}(\abs{x}-\mathcal{R})+\mathcal{R}) d\bar{\lambda}\\
&\geq U(P_{\mathcal{R}}x) + \int_0^1 (\alpha\abs{\bar{\lambda}(\abs{x}-\mathcal{R}) + \mathcal{R}}^2 - \beta) \cdot\frac{\abs{x}-\mathcal{R}}{\bar{\lambda}(\abs{x}-\mathcal{R})+\mathcal{R}} d\bar{\lambda}\\
&= U(P_{\mathcal{R}}x) + \alpha(\abs{x}-\mathcal{R})^2/2 - \beta \ln (\abs{x}/\mathcal{R}) + \alpha \mathcal{R}(\abs{x}-\mathcal{R}).
\end{align*}
It remains to show that the last two terms are nonnegative 
(note that the integral on the second-to-last line is~$\alpha\abs{x}^2/2$ in case~$\mathcal{R}=0=\beta$, so it suffices to consider~$\mathcal{R}\neq 0$). By definition of~$\mathcal{R}$, it holds for~$x\in\mathbb{R}^d\setminus B_{\mathcal{R}}$ that
\begin{equation*}
e^{\frac{\alpha}{\beta}\mathcal{R}(\abs{x}-\mathcal{R})} \geq 1+ (\alpha/\beta)\mathcal{R}(\abs{x}-\mathcal{R}) = 1+ (\abs{x}-\mathcal{R})/\mathcal{R} = \abs{x}/\mathcal{R}
\end{equation*}
which concludes after taking logarithms on both sides then multiplying by~$\beta$.
\end{proof}

In the next Proposition~\ref{cob}, the reader is referred to for example~\cite[Appendix~A]{MR4025861} for a precise definition of strong convexity on a nonconvex domain. Note that in Proposition~\ref{cob} below the assumptions are slightly different from~\eqref{set}. The subsequent Corollary~\ref{sob} deals with~\eqref{set}.
\begin{prop}\label{cob}
Suppose~$U\in C^1$. Suppose there exist~$R\geq0$,~$\bar{L}\geq\bar{m}>0$ such that~$U$ is~$\bar{m}$-strongly convex on~$\mathbb{R}^d\setminus B_R$ 
and~$\nabla U$ is~$\bar{L}$-Lipschitz. 
Assumption~\ref{A1} holds with~$c_U=0$,~$\mathcal{R}=R(1+\bar{L}/\bar{m}) + \abs{\nabla U(0)}/\bar{m}$,~$L=\bar{L}$ and~$m=\bar{m}/2$.
\end{prop}
\begin{proof}
For any~$x\in\mathbb{R}^d$ with~$\abs{x}=R$, it holds that~$\abs{\nabla U(x)}\leq \abs{\nabla U(x) - \nabla U(0)} + \abs{\nabla U(0)} \leq \bar{L}\abs{x} + \abs{\nabla U(0)} = \bar{L}R + \abs{\nabla U(0)}$. Therefore, for any~$x\in\mathbb{R}^d\setminus B_R$, it holds that
\begin{align*}
\nabla U(x)\cdot x &= (\nabla U(x) - \nabla U(P_Rx))\cdot (x-P_Rx)\abs{x}/(\abs{x}-R) + \nabla U(P_Rx)\cdot x \\
&\geq \bar{m}\abs{x-P_Rx}^2\abs{x}/(\abs{x}-R) - (\bar{L}R + \abs{\nabla U(0)})\abs{x}\\
&= \bar{m}\abs{x}(\abs{x}-R) - (\bar{L}R + \abs{\nabla U(0)})\abs{x},
\end{align*}
which implies by Young's inequality that
\begin{equation*}
\nabla U(x)\cdot x \geq \bar{m}\abs{x}^2/2 - (\bar{m}R + \bar{L}R + \abs{\nabla U(0)})^2/(2\bar{m}).
\end{equation*}
The proof concludes by Proposition~\ref{diss}.
\end{proof}
In the next Corollary~\ref{sob}, the claims in the introduction under~\eqref{set} are (together with Section~\ref{impl}) proved.
\begin{corollary}\label{sob}
Assume the settings and notations of Theorem~\ref{stu}. Assume there exist~$L\geq m>0$,~$R\geq 0$ such that~\eqref{set} holds and assume~$d\geq 2$. 
Suppose~$M=U(0)+LR^2/2$,~$c=1$ and
\begin{equation}\label{ajs}
7\sqrt{L}R(1+L/m)^{3/2}\leq \sqrt{d-1}.
\end{equation}
It holds that~$\rho\leq 26$.
\end{corollary}
\begin{proof}
We apply the second assertion in Theorem~\ref{coco}. Denote~$\kappa=L/m$. The assumption on~$M$ yields~$\bar{R}$ (as in~\eqref{Rbdef2}) satisfies
\begin{align*}
\bar{R} &=R(1+\kappa) + \sqrt{\kappa^2R^2-(2/m)U(0) + \kappa R^2 + (2/m)U(0) + \kappa R^2} \\
&= R(1+\kappa) + R\sqrt{\kappa^2 + 2\kappa}\\
& \leq 2R(1+\kappa),
\end{align*}
so that by~\eqref{ajs},
\begin{equation*}
L\bar{R}^2\kappa\leq (4/49)(d-1)
\end{equation*}
Therefore by Theorem~\ref{coco} (more specifically the consequence~\eqref{jjv}) and~\eqref{ajs}, it holds that
\begin{align*}
\rho &\leq (2e)\vee\bigg(\frac{2(2+e)(eL\bar{R}^2)^d}{ed^d\sqrt{d\pi/2}} + \frac{(1+e)}{e}\sum_{i=0}^{d-1}\bigg(\frac{4e\cdot L\bar{R}^2\kappa}{d-1}\bigg)^{d-1/2-i/2}\bigg)\\
&\leq (2e)\vee\big(2(2+e)/(e\sqrt{\pi}) + (1+e^{-1})/(1-\sqrt{16e/49})\big),
\end{align*}
which concludes the proof by a numerical bound.
\end{proof}

\section{Intractability}\label{intr}
The broad idea to show intractability appeared in~\cite[Appendix~K (Supplemental)]{NEURIPS2018_c6ede20e} (see however Remark~\ref{remt}\ref{nipt} below); their approach is based on counterexample Gaussian mixtures with two well-separated modes and different variances. 
In this section, we precisely specify when nonconvex sampling becomes intractable under quantitative assumptions. 
The assumptions considered are those of Propositions~\ref{cob} and~\ref{diss}, namely strong convexity outside a ball and distant dissipativity, with a logdensity that admits Lipschitz gradients in both cases. In the case of distant dissipativity, the counterexample will satisfy~\eqref{disseq} with~$\beta = 0$, so below this case is referred to with just dissipativity.

The first main result is Theorem~\ref{comp} in Section~\ref{bober}, in which the counterexample is close to a Gaussian mixture with two modes. 
The distance between modes that is required for the associated sampling problem to be intractable is specifically shown to scale like~$\sqrt{d}$. 
Moreover, the gradient of the logdensity will admit a Lipschitz constant that does not grow with~$d$. 
These are the main thematic differences with the example in~\cite[Appendix~K (Supplemental)]{NEURIPS2018_c6ede20e}\footnote{Their presentation is motivated by closeness to a Gaussian mixture, whereas the priority here is to exhibit a counterexample satisfying the strongest possible quantitative nonconvex assumptions.}.
Consequently, the counterexample 
satisfies the assumptions of Proposition~\ref{cob} 
with even quite stringent constants. In particular, the intractability region will match the criterion of Corollary~\ref{coco2} in the order of~$d$ as announced. More discussion on the scaling is given in Remark~\ref{remt}\ref{rem2}\ref{nipt}.

The second main Theorem~\ref{comp2}, found in Section~\ref{cocer}, concerns distant dissipativity.
In order to extend the positive results in the setting~\eqref{set} to where~\eqref{disseq} holds instead of~\eqref{set1}, one may consider for example adding a radial biasing term to the proposal logdensity~\eqref{pidef}, in the spirit of adaptive biasing force methods~\cite[Chapter~5]{MR2681239}. This (nonadapting) biasing term would drive exploration towards the flat region, which would be connected (in the sense that there is no energetic barrier) to all local modes, given a suitable condition on~$\alpha$ and~$\beta$ based on Proposition~\ref{diss} and Corollary~\ref{coco2}. Theorem~\ref{comp2} below shows that no such approach could yield polynomial complexity under such general conditions, even if~$\beta$ is restricted to~$0$.

\subsection{Strong convexity outside a ball}\label{bober}
Let~$L_0>m_0>0$ be such that~$6e^{24d^{-1}}m_0 < L_0$ and denote~$\kappa_0=L_0/m_0$. 
Let
\begin{equation}\label{c0def}
c_0=[((8/9)\sin(3\pi/16))^2 - 1/6]^{-1/2},
\end{equation}
and~$x_0\in\mathbb{R}^d$ be such that~$\abs{x_0}=c_0\sqrt{d\ln(\kappa_0)/L_0}$. 
Let~$f_1,f_2:\mathbb{R}^d\rightarrow\mathbb{R}$ be given by
\begin{equation*}
f_1(x) = \frac{m_0}{2}\abs{x}^2 + \frac{d}{2}\ln\bigg(\frac{2\pi}{m_0}\bigg),\qquad f_2(x) = \frac{L_0}{2}\abs{x-x_0}^2 + \frac{d}{2}\ln\bigg(\frac{2\pi}{L_0}\bigg),
\end{equation*}
The functions~$f_1,f_2$ are negative logdensities that will be smoothly sewn together. The assumption~$6e^{24d^{-1}}m_0 < L_0$ is made to improve the numerical constants in the main Theorem~\ref{comp} below, and it is not strictly necessary. 
Let
\begin{align}
r_d &=\sqrt{\kappa_0^{-1}(1-\kappa_0^{-1})^{-2}\abs{x_0}^2+d\ln (\kappa_0)/(L_0-m_0)}\nonumber\\
&=  \abs{x_0} \sqrt{\kappa_0^{-1} (1-\kappa_0^{-1})^{-2} + c_0^{-2}(1-\kappa_0^{-1})^{-1}}\label{rfdef}
\end{align}
with the consequence that~$B_{r_d}(x_0/(1-\kappa_0^{-1})) = \{x:f_1(x)\geq f_2(x)\}$. 
Denote
\begin{equation}\label{r0def}
\gamma:=(1-\kappa_0^{-1})^{-1}\in(1,6/5),\qquad\qquad r_0:=r_d/8.
\end{equation}
Note that by~$\kappa_0^{-1}<1/6$ and monotonicity of~$[0,1/6]\ni x\mapsto x(1-x)^{-2} + c_0^{-2}(1-x)^{-1}$, 
we have
\begin{equation}\label{rdb}
r_d/\abs{x_0} \in [c_0^{-1},\gamma(1/6 + c_0^{-2})^{1/2}].
\end{equation}
Let~$g:\mathbb{R}^d\rightarrow[0,1]$ be given by
\begin{equation}\label{gdef}
g(x) = \begin{cases}
1 & \textrm{if } x\in \mathbb{R}^d\setminus B_{r_d+r_0}(\gamma x_0),\\
\int_{-1}^{\abs{x-P_{B_{r_d-r_0}(\gamma x_0)}x}/r_0-1} 
\varphi(y)
dy & \textrm{if } x\in B_{r_d+r_0}(\gamma x_0)\setminus B_{r_d-r_0}(\gamma x_0), \\
0 & \textrm{if } x\in B_{r_d-r_0}(\gamma x_0),
\end{cases}
\end{equation}
where recall from the notation section that~$\varphi$ denotes the standard mollifier and~$P_{B_R(y)}$ denotes the projection to the ball~$B_R(y)$. 
Let~$f_3:\mathbb{R}^d\rightarrow\mathbb{R}$ be given by
\begin{equation}\label{f3def}
f_3 = gf_1 + (1-g)f_2. 
\end{equation}
The following Lemma~\ref{f2mass} 
estimates the mass contributed by~$f_2$ 
in the combined landscape~$f_3$. 
\begin{lemma}\label{f2mass}
It holds that
\begin{equation}\label{f2eq}
\int_{B_{r_d-r_0}(\gamma x_0)} e^{-f_3} \geq \frac{1}{4}\int_{\mathbb{R}^d} e^{-f_3}.
\end{equation}
\end{lemma}
\begin{proof}
Denote~$a=\gamma\kappa_0^{-1}\abs{x_0}$,~$b=(\sqrt{d+1}+\sqrt{2\ln2})/\sqrt{L_0}$. 
Firstly, since~$\kappa_0$ satisfies~$\kappa_0\geq 6e^{\frac{24}{d}}$, 
it holds that~$\ln\kappa_0\geq \ln 6 + 24/d$ and so
\begin{align}
(a+b)^2 &= \bigg|\begin{pmatrix}a\kappa_0^{1/2}\\(\gamma d\ln\kappa_0)^{1/2}L_0^{-1/2}\end{pmatrix}  \cdot\begin{pmatrix} \kappa_0^{-1/2}\\ b(\gamma d\ln\kappa_0)^{-1/2}L_0^{1/2}
\end{pmatrix}\bigg|^2\nonumber\\
&\leq (a^2\kappa_0 + L_0^{-1}\gamma d\ln\kappa_0)\cdot ( \kappa_0^{-1} + b^2L_0(\gamma d \ln \kappa_0)^{-1})\nonumber\\
&\leq r_d^2\cdot ( \kappa_0^{-1} + (1-\kappa_0^{-1}) [(\sqrt{d+1}+\sqrt{2\ln2})^2(24+d\ln 6)^{-1}]),\label{rdc}
\end{align}
where we have used the definitions~\eqref{rfdef} and~\eqref{r0def} of~$r_d^2,\gamma$. On the right-hand side, the expression in the square brackets satisfies
\begin{align*}
(\sqrt{d+1}+\sqrt{2\ln2})^2(24+d\ln 6)^{-1} &= (d+1+2\ln2 + 2\sqrt{(d+1)\cdot 2\ln2})(24+d\ln 6)^{-1}\\
&\leq ((11/10)(d+1)+11\cdot 2\ln2 )(24+d\ln 6)^{-1}\\
&\leq 7/10
\end{align*}
so that together with the assumption~$\kappa_0\geq 6$ and the monotonicity of the right-hand side of~\eqref{rdc} w.r.t.~$\kappa_0$, it holds that
\begin{equation*}
(a+b)^2\leq r_d^2\cdot (1/6 + (5/6)7/10)\leq (49/64) r_d^2,
\end{equation*}
which is, after taking square roots and by definition~$r_0=r_d/8$,
\begin{equation}\label{cda}
r_d-r_0\geq \gamma \kappa_0^{-1}\abs{x_0} + \sqrt{(d+1)/L_0} + \sqrt{2\ln (2)/L_0}. 
\end{equation}
Moreover, by a spherical coordinate transform, it holds that
\begin{align*}
\int_{\mathbb{R}^d} \abs{x-x_0}e^{-f_2(x)}dx &= \int_0^{\infty} \!\! r^d e^{-L_0r^2/2} dr \bigg( \int_0^{\infty} \!\! r^{d-1} e^{-L_0r^2/2} dr \bigg)^{\!\!-1} \!\!= \frac{\sqrt{2}\Gamma((d+1)/2)}{\sqrt{L_0}\Gamma(d/2)}.
\end{align*}
Therefore, by Theorems~5.2,~5.3 in~\cite{MR1849347} with~$F=\abs{\cdot}$ and~$\mu$ equal to the~$L_0^{-1}I_d$-variance Gaussian, it holds that
\begin{align*}
\int_{B_{r_d-r_0}(\gamma x_0)} e^{-f_2} &\geq \int_{B_{r_d-r_0-\gamma\kappa_0^{-1} \abs{x_0}}(x_0)} e^{-f_2} \\
&\geq 1- \exp\bigg(\!\!-\frac{L_0}{2}\bigg(r_d-r_0-\gamma\kappa_0^{-1} \abs{x_0}- \frac{ \sqrt{2}\Gamma((d+1)/2)}{\sqrt{L_0}\Gamma(d/2)}\bigg)^{\!\!2}\,\bigg),
\end{align*}
Consequently, by the definitions~\eqref{gdef},~\eqref{f3def} of~$g,f_3$ and given~\eqref{cda} together with Gautschi's inequality, the integral on the left-hand side of~\eqref{f2eq} satisfies
\begin{equation*}
\int_{B_{r_d-r_0}(\gamma x_0)} e^{-f_3} = \int_{B_{r_d-r_0}(\gamma x_0)} e^{-f_2} \geq \frac{1}{2}.
\end{equation*}
On the other hand, by Young's inequality, the total mass satisfies
\begin{align*}
\int_{\mathbb{R}^d}e^{-f_3} = \int_{\mathbb{R}^d}e^{-gf_1 -(1-g)f_2} \leq \int_{\mathbb{R}^d}(e^{-f_1} +  e^{-f_2}) = 2,
\end{align*}
which concludes the proof.
\end{proof}

Next, it is shown that~$\nabla f_3$ admits a well-behaved (uniform in~$d$) global Lipschitz constant.
\begin{lemma}\label{f3gc}
For any~$x\in\mathbb{R}^d$, the operator norm of the Hessian~$D^2 f_3(x)$ satisfies~$\abs{D^2 f_3(x)} \leq (264 + 95c_0 + 21c_0^2)L_0 \leq 878L_0$.
\end{lemma}
\begin{proof}
By definition~\eqref{f3def} of~$f_3$, the Hessian~$D^2 f_3$ satisfies
\begin{align}
D^2 f_3 &= D^2g f_1 +  \nabla f_1(\nabla g)^{\top} + \nabla g (\nabla f_1)^{\top}  + gD^2f_1\nonumber\\
&\quad - D^2g f_2 - \nabla f_2 (\nabla g)^{\top} - \nabla g (\nabla f_2)^{\top} + (1-g)D^2 f_2.\label{d2f3}
\end{align}
To estimate the Hessian~$D^2g$, by definition~\eqref{gdef} of~$g$, note it holds for~$x\in B_{r_d+r_0}(\gamma x_0)\setminus B_{r_d-r_0}(\gamma x_0)$ that
\begin{equation}\label{gradg}
\nabla g(x) = \varphi(\abs{x-P_{B_{r_d-r_0}(\gamma x_0)}x}/r_0 - 1) r_0^{-1} (x-\gamma x_0)/\abs{x-\gamma x_0},
\end{equation}
so that
\begin{align*}
D^2 g(x) &= \partial \varphi (\abs{x-P_{B_{r_d-r_0}(\gamma x_0)}x}/r_0 - 1) r_0^{-2} (x- \gamma x_0)^{\otimes 2}/\abs{x-\gamma x_0}^2\\
&\quad+ \varphi (\abs{x-P_{B_{r_d-r_0}(\gamma x_0)}x}/r_0 - 1)r_0^{-1} \bigg(\frac{I_d}{\abs{x-\gamma x_0}} - \frac{(x-\gamma x_0)^{\otimes 2}}{\abs{x-\gamma x_0}^3}\bigg),
\end{align*}
where~$\bar{v}^{\otimes 2} = \bar{v}\bar{v}^{\top}\in\mathbb{R}^{d\times d}$ for vectors~$\bar{v}\in\mathbb{R}^d$. 
Therefore, by numerically approximating~$\max_x\abs{\varphi(x)}\leq 21/25,\max_x\abs{\partial\varphi(x)}\leq 9/5$, the Hessian~$D^2g(x)$ for~$x\in B_{r_d+r_0}(\gamma x_0)\setminus B_{r_d-r_0}(\gamma x_0)$ satisfies
\begin{align*}
\abs{D^2g(x)}
&\leq (9/5)r_0^{-2} + (42/25)r_0^{-1}\abs{x-\gamma x_0}^{-1}\\
&\leq (9/5)r_0^{-2} + (42/25)r_0^{-1}(r_d-r_0)^{-1},
\end{align*}
so that
\begin{align}
&\abs{D^2g(x)(f_1(x)-f_2(x))} \nonumber\\
&\quad\leq ((9/5)r_0^{-2} + (42/25)r_0^{-1}(r_d-r_0)^{-1}) \cdot(m_0(r_d+r_0+\gamma\abs{x_0})^2/2 \nonumber\\
&\qquad + L_0(r_d+r_0+(\gamma-1)\abs{x_0})^2/2 + d\abs{\ln(2\pi/m_0)-\ln(2\pi/L_0)}/2)\nonumber\\
&\quad\leq 64(51/25)r_d^{-2} \cdot((1/12)L_0((9/8)r_d+\gamma\abs{x_0})^2 + L_0((9/8)r_d + (\gamma-1)\abs{x_0})^2 \nonumber\\
&\qquad+ d\ln(\kappa_0)/2).
\label{d2gf}
\end{align}
By~\eqref{rdb}, we have~$\abs{x_0} \leq c_0r_d$, and by~\eqref{rfdef}, we have~$d\ln(\kappa_0)\leq r_d^2(L_0-m_0)$, therefore, together with~\eqref{r0def}, 
inequality~\eqref{d2gf} implies
\begin{equation}\label{d2gfs}
\abs{D^2g(x)(f_1(x)-f_2(x))}\leq (245 +89c_0 + 21c_0^2)L_0
\end{equation}
For the terms with first order derivatives in~\eqref{d2f3}, it is easy to derive corresponding bounds given~\eqref{gradg}. More specifically, it holds for~$x\in B_{r_d+r_0}(\gamma x_0)\setminus B_{r_d-r_0}(\gamma x_0)$ that
\begin{align*}
&\abs{(\nabla f_1 (\nabla g)^{\top} + \nabla g(\nabla f_1)^{\top} - \nabla f_2 (\nabla g)^{\top} - \nabla g(\nabla f_2)^{\top})(x)} \\
&\quad\leq 2(21/25)r_0^{-1}(m_0(r_d+r_0+\gamma\abs{x_0}) + L_0(r_d + r_0 + (\gamma-1)\abs{x_0}))\\
&\quad\leq (42/25)r_0^{-1}((7/6)L_0\cdot (9/8)r_d+((1/6)L_0\gamma+L_0(\gamma-1))\abs{x_0})\\
&\quad\leq 8\cdot(42/25)L_0((7/6)\cdot(9/8) +(2/5)c_0 )\\
&\quad\leq (18 +  6c_0)L_0.
\end{align*}
Gathering with~\eqref{d2gfs},~$D^2f_1= m_0I_d$,~$D^2f_2=L_0I_d$ and substituting these into a bound for~\eqref{d2f3} concludes the proof.
\end{proof}

The following Theorem~\ref{comp} shows that for any algorithm (represented by~$\Phi_j$, see Remark~\ref{remt}\ref{remt1} below) and any reasonable form of estimator (represented by~$\hat{\theta}_j$, see again Remark~\ref{remt}\ref{remt1} below), there is a counterexample target density and integrand pair such that if the number of iterations is less than exponential-in-$d$, then the resulting estimator of the integral with respect to the target measure is poor.

\begin{theorem}\label{comp}
Assume~$d>1$. Let~$\bar{x}_0$ be an~$\mathbb{R}^d$-valued r.v. and let~$N\in\mathbb{N}$. 
For any~$j\in\mathbb{N}$, let~$\Phi_j: \Omega\times(\mathbb{R}^d)^j\times \mathbb{R}^j \times (\mathbb{R}^d)^j\times \mathbb{R}^j\rightarrow\mathbb{R}^d$ 
and~$\hat{\theta}_j: \Omega\times(\mathbb{R}^d)^{j+1}\times \mathbb{R}^{j+1} \times (\mathbb{R}^d)^{j+1}\times \mathbb{R}^{j+1}\rightarrow\mathbb{R}$ 
be measurable random functions such that there exists a constant~$\abs{\hat{\theta}_j}\in[1,\infty)$ satisfying a.s. that
\begin{equation}\label{theb}
\hat{\theta}_j(\bar{x}_0,\dots \bar{x}_j,y_0,\dots,y_j,\bar{y}_0,\dots,\bar{y}_j,z_0,\dots z_j)  \leq \abs{\hat{\theta}_j}\max_k\abs{z_k}
\end{equation}
for all~$\bar{x}_k,\bar{y}_k\in\mathbb{R}^d$,~$y_k,z_k\in\mathbb{R}$ with~$k\in[0,j]\cap\mathbb{N}$. 
For any~$j\in\mathbb{N}$,~$f\in C^1(\mathbb{R}^d)$ and~$\theta:\mathbb{R}^d\rightarrow\mathbb{R}$, let~$\hat{x}_j^{f,\theta} \in (\mathbb{R}^d)^{j+1}\times \mathbb{R}^{j+1} \times (\mathbb{R}^d)^{j+1}\times \mathbb{R}^{j+1}$ denote
\begin{align*}
\hat{x}_j^{f,\theta} &= (x_0^{f,\theta},\dots,x_j^{f,\theta},f(x_0^{f,\theta}),\dots,f(x_j^{f,\theta}),\\
&\qquad\nabla f(x_0^{f,\theta}),\dots,\nabla f(x_j^{f,\theta}),\theta(x_0^{f,\theta}),\dots,\theta(x_j^{f,\theta})),
\end{align*}
where for any~$j\in\mathbb{N}$,~$x_j^{f,\theta}$ is an~$\mathbb{R}^d$-valued r.v. 
such that~$x_{j+1}^{f,\theta} = \Phi_{j+1}(\hat{x}_j^{f,\theta})$ and~$x_0^{f,\theta}=\bar{x}_0$. 
There exist bounded~$\phi\in C^{\infty}(\mathbb{R}^d)$ with bounded derivatives of all orders and~$U\in C^\infty(\mathbb{R}^d)$ satisfying~$\nabla U(0)=0$, 
\begin{align*}
u^{\top}D^2U(x)u&\geq m_0\abs{u}^2& & 
\forall u\in\mathbb{R}^d,x\in\mathbb{R}^d\setminus B_{7\sqrt{d\ln(\kappa_0)/L_0}}\,,\\
\abs{D^2 U(x)}&\leq 878 L_0 & & \forall x\in\mathbb{R}^d
\end{align*}
such that if 
\begin{equation}\label{Nin2}
N < \frac{(\sin(3\pi/8))^{-d}}{5 (d+2)^{1/2}\abs{\hat{\theta}_N}}-1,
\end{equation}
then it holds that
\begin{equation}\label{compeq}
\bigg|\mathbb{E}\bigg[\hat{\theta}_N(\hat{x}_N^{U,\phi}) - \int_{\mathbb{R}^d}\frac{\phi e^{-U}}{Z}\bigg]\bigg|\geq \frac{\max_x\abs{\phi(x)}}{8}\vee\bigg(\bigg(\frac{d\ln (\kappa_0)}{L_0}\bigg)^{\!\frac{1}{2}}\frac{\max_x\abs{\nabla \phi(x)}}{64}\bigg).
\end{equation}
\end{theorem}
\begin{remark}\label{remt}
\begin{enumerate}[label=(\roman*)]
\item \label{remt1}
The functions~$\Phi_j$ represent the~$j^{\textrm{th}}$-steps in any stochastic algorithm that may be applied to a sampling problem associated to the negative logdensity~$f$. 
Each~$\Phi_j$ is used to produce a point~$x_j^{f,\theta}$, at which~$f,\nabla f,\theta$ may possibly be evaluated to obtain~$(x_{j+i}^{f,\theta})_{i> 0}$. The class of algorithms represented by~$(\Phi_j)_j$ includes for example all of those considered in~\cite{syed2024} (parallel tempering, sequential Monte Carlo, annealed IS). The tail-matching procedure described in the present article is no exception. 
The functions~$\hat{\theta}_j$ represent the form of any estimators for integrals of interest with respect to the target distribution, given~$\hat{x}_j^{f,\theta}$. Indeed,~$\hat{\theta}_j$ is typically some (weighted) average of evaluations of the integrand at points~$(x_k^{f,\theta})_{k\in[0,j]\cap\mathbb{N}}$. In particular, 
the constant~$\abs{\hat{\theta}_j}$ in~\eqref{theb} satisfies~$\abs{\hat{\theta}_j}=1$ for all~$j$ and it suffices to write~$\hat{\theta}_j = \hat{\theta}_j(y_0,\dots,y_j,z_0,\dots,z_j)$ on the left-hand side of~\eqref{theb}. 
Finally, the notation~$\hat{x}_j^{f,\theta}$ represents all of the information available from the problem given the past trajectory~$(x_k^{f,\theta})_{k\in[0,j]\cap\mathbb{N}}$. 
\item \label{rem2} Theorem~\ref{comp} shows that for any large enough (asymptotic) condition number~$\kappa=L/m$, there is a counterexample logdensity in every dimension~$d$ with this condition number and 
satisfying~\eqref{set} 
with~$\sqrt{L}R = 7\sqrt{878}\cdot\sqrt{d\ln(878^{-1}L/m))}\leq 208\sqrt{d\ln(1+\kappa)}$ such that the corresponding measure cannot be sampled accurately in polynomial time. By Corollary~\ref{sob} and the results of Section~\ref{impl}, this matches the order of the tractability bound for the tail-matching importance sampler up to a polynomial factor in the condition number, see Remark~\ref{remo}\ref{rem3} and Remark~\ref{jjq}. 
\item \label{nipt} The general intuition in the proof of Theorem~\ref{comp} is already present in Theorem~K.1 of~\cite[Supplemental]{NEURIPS2018_c6ede20e}; the idea is to use that the number of directions where important modes may appear increases exponentially in~$d$. However, 
the conclusion here is more general, accommodating self-normalized IS estimators, and special considerations are made for the Lipschitz and strong convexity constants. 
In particular, the tractability/intractability regions are identified for~$\sqrt{L}\mathcal{R}$ in terms of~$d,L/m$, which is slightly more subtle than the fixed~$m,L$ particular case (see also item~\ref{rem2} above and the discussion at the beginning of Section~\ref{intr}).
\item 
The proof of Theorem~\ref{comp} is clearly not limited to algorithms using only zeroth and first order (gradient) information. Moreover, it is clear from the computations in this section that the result is also relevant for cases where Lipschitz conditions on the Hessian of the target logdensity are assumed.  These conditions appear in parts of the literature to improve theoretical bounds. In addition, estimators based on control variates for variance reduction are not considered, but similar results can be expected to hold.
\item Theorem~\ref{comp} is a statement about the limitations to the positive conclusions that can be drawn from general quantitative nonconvexity assumptions. For any particular nonlogconcave target and test function, it may well be the case that there exist efficient sampling algorithms, which is discussed in the next point~\ref{last}. However if the only a priori information available about them is that they satisfy such general assumptions and point evaluations of the target are otherwise black-box operations, then Theorem~\ref{comp} applies with effect.
\item \label{last} The starting point~$\bar{x}_0$, the algorithm procedure~$(\Phi_j)_j$ and the form~$\hat{\theta}_N$ of the estimator are fixed before~$U$ and~$\phi$. 
If there is accurate a priori information about a given target~$U$ or~$\phi$ that is used to choose~$\bar{x}_0$,~$(\Phi_j)_j$ or~$\hat{\theta}_N$, then Theorem~\ref{comp} does not apply. 
The idea to use a priori information to speed up nonconvex sampling can be found in for example~\cite{MR4152629}. In that work, it is the goal of a mode-finding precomputation to locate local optima of the target logdensity, which are then used in the main sampling algorithm to facilitate fast switching between the modes. However, an implicit assumption is made that all of the modes have been found during this initial stage. Their full implementation including the mode-finding component also suffers from the negative result above. 
Use of a priori information is also common in molecular dynamics, in the form of \textit{reaction coordinates}~\cite{MR2681239}, which in some cases may be formed from expert knowledge of the physical system associated with the sampling problem. Some recent approaches~\cite{zgcref,10.1063/5.0151053} combine a mix of expert knowledge and machine learning to obtain this information. On the other hand, if reaction coordinates are to be concocted purely by point evaluations of~$U,\phi$ and their derivatives~\cite{zineb1,pmlr-v108-p}, then Theorem~\ref{comp} above applies for considerations on theoretical guarantees. See also~\cite{pmlr-v291-koehler25a,lee2025s,monmarche2026g} for works making use of such additional information.
\end{enumerate}
\end{remark}
\begin{proof}[Proof of Theorem~\ref{comp}]
Assume~\eqref{Nin2}. 
Recall the notation~$\varphi$ for the standard mollifier and note~$\max_x{\varphi(x)}\leq 1$. 
For any~$x\in\mathbb{R}^d$, let~$h_x:\mathbb{R}^d\rightarrow[0,1]$ be given by
\begin{equation*}
h_x(z) = \begin{cases}
1 & \textrm{if }z\in\mathbb{R}^d\setminus B_{r_d+r_0}(\gamma x),\\
\int_{-1}^{\abs{z-P_{B_{r_d-r_0}(\gamma x)}z}/r_0-1} \!\varphi(y) dy &\textrm{if } z\in B_{r_d+r_0}(\gamma x) \setminus B_{r_d-r_0}(\gamma x),\\
0 & \textrm{if }z\in B_{r_d-r_0}(\gamma x)
\end{cases}
\end{equation*}
with the consequence that~$\max_z\abs{h_x(z)}= 1$ and~$\max_z\abs{\nabla h_x(z)}\leq r_0^{-1} = 8/r_d\leq 8\sqrt{L_0/(d\ln(\kappa_0))}$ by~\eqref{rdb}. 
For any~$x\in\mathbb{R}^d$, let~$f_x$ denote~$f_3$ as in~\eqref{f3def} with~$x_0=x$. 
Suppose for contradiction that for any~$x_0\in\mathbb{R}^d$ with~$\abs{x_0}=c_0\sqrt{d\ln(\kappa_0)/L_0}$, inequality~\eqref{compeq} with~$U=f_{x_0}$ and~$\phi=1-h_{x_0}$ does not hold, in other words, it holds that
\begin{equation}\label{compeqneg}
\bigg|\mathbb{E}\bigg[\hat{\theta}_N(\hat{x}_N^{f_{x_0},1-h_{x_0}}) - \int_{\mathbb{R}^d}(1-h_{x_0}) e^{-f_{x_0}}\bigg(\int_{\mathbb{R}^d} e^{-f_{x_0}}\bigg)^{-1}\bigg]\bigg| < \frac{1}{8}.
\end{equation}
For any~$x_0$ with~$\abs{x_0}=c_0\sqrt{d\ln(\kappa_0)/L_0}$, Lemma~\ref{f2mass} implies 
\begin{equation*}
\int_{\mathbb{R}^d}(1-h_{x_0}) e^{-f_{x_0}}\bigg(\int_{\mathbb{R}^d}e^{-f_{x_0}} \bigg)^{-1} \geq \frac{1}{4}.
\end{equation*}
Therefore, it holds for any such~$x_0$ that~$\mathbb{E}[\hat{\theta}_N(\hat{x}_N^{f_{x_0},1-h_{x_0}})] > 1/8$. 
If the event~$\mathcal{K}\subset\Omega$ defined by
\begin{equation*}
\mathcal{K}=\cap_{i=0}^N\{x_i^{f_{x_0},1-h_{x_0}} \notin \textrm{int}(B_{r_d+r_0}(\gamma x_0))\}
\end{equation*}
satisfies the inequality~$\mathbb{P}(\Omega\setminus\mathcal{K}) \leq 1/(8\abs{\hat{\theta}_N})$, then~\eqref{theb} implies
\begin{align*}
\mathbb{E}[\hat{\theta}_N(\hat{x}_N^{f_{x_0},1-h_{x_0}})] &= \mathbb{E}[\mathds{1}_{\mathcal{K}}\hat{\theta}_N(\hat{x}_N^{f_{x_0},1-h_{x_0}})] + \mathbb{E}[\mathds{1}_{\Omega\setminus\mathcal{K}}\hat{\theta}_N(\hat{x}_N^{f_{x_0},1-h_{x_0}})] \\
&\leq \mathbb{E}[\mathds{1}_{\Omega\setminus\mathcal{K}}\abs{\hat{\theta}_N}\textstyle\max_x(1-h_{x_0}(x))]\\
&\leq 1/8,
\end{align*}
which is a contradiction. Therefore~$\mathcal{K}$ satisfies~$\mathbb{P}(\mathcal{K})< 1-1/(8\abs{\hat{\theta}_N})$ for any such~$x_0$. 
Equivalently, it holds that
\begin{equation}\label{ppo}
\mathbb{P}(\exists i\in[0,N]\cap\mathbb{N}: x_i^{f_{x_0},1-h_{x_0}}\in \textrm{int}(B_{r_d+r_0}(\gamma x_0))) \geq 1/(8\abs{\hat{\theta}_N}).
\end{equation}
Let~$\mathcal{G}_0\subset \mathbb{S}^{d-1}=\{x\in\mathbb{R}^d:\abs{x} = 1\}$ be a finite set such that for any~$\bar{x},\bar{y}\in\mathcal{G}_0$ with~$\bar{x}\neq\bar{y}$ it holds that~$\bar{x}\cdot\bar{y} <\cos(3\pi/8)$ and such that it has size~$\abs{\mathcal{G}_0} = S(3\pi/8)$, where~$S:(0,\pi/2)\rightarrow\mathbb{N}$ is given for any~$\bar{\alpha}\in(0,\pi/2)$ by
\begin{equation}\label{jga}
S(\bar{\alpha}) = \max \{\abs{\bar{\mathcal{G}}}: \bar{x}\cdot \bar{y} < \cos(\bar{\alpha}) \quad \forall \bar{x},\bar{y}\in \bar{\mathcal{G}}\subset\mathbb{S}^{d-1}, \bar{x}\neq \bar{y} \}.
\end{equation}
By the lower bound in~\cite[equation~(24)]{MR180417}, 
together with Gautschi's inequality, it holds that
\begin{align}
\abs{\mathcal{G}_0} &\geq \frac{d\sqrt{\pi}}{d-1}\cdot\frac{\Gamma((d+1)/2)}{\Gamma(d/2+1)}\bigg(\int_0^{3\pi/8}(\sin(\bar{\varphi}))^{d-2}d\bar{\varphi}\bigg)^{-1}\nonumber\\
&\geq d\sqrt{2\pi}((d-1)\sqrt{d+2})^{-1}(3\pi/8)^{-1}(\sin(3\pi/8))^{2-d}.\label{gsi}
\end{align}
Let~$\mathcal{G}=\{c_0g\sqrt{d\ln(\kappa_0)/L_0}:g\in \mathcal{G}_0\}$, 
The elements in~$\mathcal{G}$ 
will be considered as values for~$x_0$ in the above. For any~$i\in[1,\abs{\mathcal{G}}]\cap\mathbb{N}$, let~$y_i\in\mathcal{G}$, such that~$y_i\neq y_j$ for~$i\neq j$. For any~$i,j$ with~$i\neq j$, it holds by definitions that
\begin{equation*}
\abs{\gamma y_i-\gamma y_j}\geq 2\gamma\sin(3\pi/16)\cdot c_0\sqrt{d\ln(\kappa_0)/L_0}.
\end{equation*}
Moreover,~\eqref{rdb} implies~$r_d+r_0 = (9/8)r_d \leq (9/8)\gamma (1/6+c_0^{-2})^{1/2} c_0\sqrt{d\ln(\kappa_0)/L_0}$. 
Therefore by definition~\eqref{c0def} of~$c_0$, 
it holds for~$i\neq j$ that~$\textrm{int}(B_{r_d+r_0}(\gamma y_i))\cap \textrm{int}(B_{r_d+r_0}(\gamma y_j)) = \emptyset$. 
For any~$j\in[1,\abs{\mathcal{G}}]\cap\mathbb{N}$, define
\begin{equation*}
 A_j:=\{\exists i\in[0,N]\cap\mathbb{N}:x_i^{f_{y_j},1-h_{y_j}}\in\textrm{int}(B_{r_d+r_0}(\gamma y_j))\} \subset \Omega.
\end{equation*}
By~\eqref{ppo},~$A_j$ satisfies
\begin{equation}\label{Pa}
\mathbb{P}(A_j) \geq 1/(8\abs{\hat{\theta}_N}) \qquad\forall j\in[1,\abs{\mathcal{G}}]\cap\mathbb{N}.
\end{equation}
On the other hand, for any~$M\in\mathbb{N}$ with~$M\in(N,\abs{\mathcal{G}})\neq\emptyset$ and~$j_0,\dots,j_M\in[1,\abs{\mathcal{G}}]\cap\mathbb{N}$ such that~$j_i\neq j_k$ for all~$i,k$, if~$\omega\in \cap_{i=0}^M A_{j_i} $ holds, then for any~$k\in[0,M]\cap\mathbb{N}$ there exists~$i\in[0,N]\cap\mathbb{N}$ satisfying~$x_i^{f_1,0}(\omega) \in \textrm{int}(B_{r_d+r_0}(\gamma y_{j_k}))$ by definition, which contradicts~$M>N$. Therefore for any such~$M$ and~$j_0,\dots,j_M$, it holds that~$\cap_{i=0}^M A_{j_i} = \emptyset$. 
In particular, denoting~$[\abs{\mathcal{G}}]_{-j}=([1,\abs{\mathcal{G}}]\cap\mathbb{N})\setminus\{j\}$ for any~$j\in[1,\abs{\mathcal{G}}]\cap\mathbb{N}$, in the expansion
\begin{equation}\label{enp}
\sum_{j=1}^{\abs{\mathcal{G}}} \mathbb{P}(A_j) =\sum_{j=1}^{\abs{\mathcal{G}}}\sum_{(\bar{A}_k)_k\in\prod_{k\in[\abs{\mathcal{G}}]_{-j}}\{A_k,\Omega\setminus A_k\}}\mathbb{P}\bigg(A_j\cap\bigcap_{k\in[\abs{\mathcal{G}}]_{-j}} \bar{A}_{k}\bigg),
\end{equation}
no nonzero summand on the right-hand side appears in the double sum expansion (strictly) more than~$N+1$ times.
Together with~\eqref{Pa}, this implies
\begin{align}
\frac{\abs{\mathcal{G}}}{8\abs{\hat{\theta}_N}} &\leq \sum_{j=1}^{\abs{\mathcal{G}}} \mathbb{P}(A_j) \nonumber\\
&\leq (N+1)\sum_{(\bar{A}_k)_k\in\prod_{k\in[1,\abs{\mathcal{G}}]\cap\mathbb{N}}\{A_k,\Omega\setminus A_k\}} \mathbb{P}\bigg(\bigcap_{k\in[1,\abs{\mathcal{G}}]\cap\mathbb{N}} \bar{A}_k\bigg)\nonumber\\
&=N+1,\label{enq}
\end{align}
which contradicts~\eqref{Nin2} with~\eqref{gsi}.
\end{proof}

\subsection{Dissipativity}\label{cocer}
Recall that~$\varphi$ denotes the standard mollifier. 
Let~$H_0:\mathbb{R}\rightarrow [0,1]$ be the shifted Heaviside function given by
\begin{equation*}
H_0(x)=\begin{cases}
0 &\textrm{if } x \leq (3\sqrt{3}+\sqrt{2})/8 \\
1 &\textrm{if } x >(3\sqrt{3}+\sqrt{2})/8.
\end{cases}
\end{equation*}
Let~$H:\mathbb{R}\rightarrow [0,1]$ be given by~$H(x) = \big(\frac{8}{\sqrt{3}-\sqrt{2}}\big)\int_{\mathbb{R}} \varphi \big((x-y)/\big(\frac{\sqrt{3}-\sqrt{2}}{8}\big)\big) H_0(y) dy$, satisfying in particular
\begin{equation}\label{Hdef}
H(x) = \begin{cases}
0 &\textrm{if } x\leq (\sqrt{3}+\sqrt{2})/4\\
1 &\textrm{if } x \geq \sqrt{3}/2.
\end{cases}
\end{equation}
Let~$m_1,L_1>0$ be such that~$16m_1\leq L_1$ and denote~$\kappa_1=L_1/m_1$. Let~$z\in\mathbb{R}^d$ be such that~$\abs{z}=1$. 
Let the negative logdensity~$f_4:\mathbb{R}^d\rightarrow [0,\infty)$ be given by~$f_4(0)=0$ and for any~$x\neq 0$ by
\begin{equation}\label{f4def}
f_4(x) = \big(m_1H(z\cdot x/\abs{x}) + L_1(1-H(z\cdot x/\abs{x}))\big) \abs{x}^2/2.
\end{equation}
Just as in the previous Section~\ref{bober}, the two things to check before the main Theorem~\ref{comp2} are that the direction in~$z$ has plenty of weight in the density~$e^{-f_4}$ and that gradient~$\nabla f_4$ is Lipschitz with a well-controlled constant. This is the content of the next Lemmata~\ref{f4mass} and~\ref{f4gc} respectively.

For simplicity, the~$d>2$ case is considered, but similar results hold in the dimensions~$d=1,2$. 
\begin{lemma}\label{f4mass}
For~$d>2$, 
it holds that
\begin{equation*}
\int_{\{x\neq 0 : z\cdot x/\abs{x} \geq (\sqrt{3}+\sqrt{2})/4\}} e^{-f_4} \geq \frac{1}{2} \int_{\mathbb{R}^d} e^{-f_4}.
\end{equation*}
\end{lemma}
\begin{proof}
By a spherical coordinate transform on Euclidean space with~$z\in\mathbb{R}^d$ as the first basis vector, 
it holds by~\eqref{Hdef} that
\begin{align}\label{jdq}
&\int_{\{x\in\mathbb{R}^d\setminus \{0\} : z\cdot x/\abs{x} \geq \sqrt{3}/2\}} e^{-f_4} \nonumber\\
&\quad= \frac{2\pi^{\frac{d-1}{2}}}{\Gamma((d-1)/2)} \int_{0}^{\infty}\int_0^{\frac{\pi}{6}} r^{d-1}\sin^{d-2}(\theta) e^{-m_1r^2/2} d\theta dr.
\end{align}
The angular integral on the right-hand side of~\eqref{jdq} may be bounded as
\begin{equation*}
\int_0^{\frac{\pi}{6}} \sin^{d-2}(\theta)  d\theta \geq \int_{\frac{\pi}{12}}^{\frac{\pi}{6}} \sin^{d-2}(\theta)  d\theta \geq \frac{\pi}{12\cdot 4^{d-2}} >  \frac{\pi}{4^d}.
\end{equation*}
The radial integral on the right-hand side of~\eqref{jdq} may be bounded by Gautschi's inequality as
\begin{equation*}
\int_0^{\infty} r^{d-1} e^{-m_1r^2/2} dr 
= \frac{2^{\frac{d}{2}-1}}{m_1^{d/2}} \Gamma\bigg(\frac{d}{2}\bigg) > \frac{2^{\frac{d}{2}-1}}{m_1^{d/2}} \bigg(\frac{d}{2}-1\bigg)^{\!\frac{1}{2}}\Gamma\bigg(\frac{d-1}{2}\bigg).
\end{equation*}
Denote~$\mathcal{C} = \{x\in\mathbb{R}^d\setminus \{0\}: z\cdot x/\abs{x} \geq (\sqrt{3}+\sqrt{2})/4\}$. 
Substituting the last two inequalities back into~\eqref{jdq} yields
\begin{equation*}
\int_{\mathcal{C}} e^{-f_4} \geq \sqrt{\pi}\bigg(\frac{\pi}{8m_1}\bigg)^{\!\frac{d}{2}}\bigg(\frac{d}{2}-1\bigg)^{\!\frac{1}{2}}.
\end{equation*}
On the other hand, the total mass satisfies~$\int_{\mathbb{R}^d} e^{-f_4} = \int_{\mathbb{R}^d\setminus \mathcal{C} } e^{-f_4} + \int_{\mathcal{C}} e^{-f_4}$, 
where the first integral on the right-hand side may be bounded by definition of~$f_4$ as
\begin{equation*}
\int_{\mathbb{R}^d\setminus \mathcal{C}} e^{-f_4} 
= \int_{\mathbb{R}^d\setminus\mathcal{C} } e^{-L_1\abs{x}^2/2}dx 
\leq \int_{\mathbb{R}^d} e^{-L_1\abs{x}^2/2}dx = \bigg(\frac{2\pi}{L_1}\bigg)^{\!\frac{d}{2}}.
\end{equation*}
Gathering the last inequalities, it holds that
\begin{align*}
\int_{\mathcal{C}} e^{-f_4} \bigg(\int_{\mathbb{R}^d} e^{-f_4}\bigg)^{\!-1} 
&= \bigg(1+\int_{\mathbb{R}^d\setminus \mathcal{C}} e^{-f_4} \bigg( \int_{\mathcal{C} } e^{-f_4}\bigg)^{\!-1}\,\bigg)^{\!-1}\\
&\geq \bigg(1+ \frac{1}{\sqrt{\pi(d/2-1)}}\cdot\bigg(\frac{16}{\kappa_1}\bigg)^{\!\frac{d}{2}}\,\bigg)^{\!-1},
\end{align*}
which concludes by the original assumption~$16m_1\leq L_1$.
\end{proof}
\begin{lemma}\label{f4gc}
For any~$x,y\in\mathbb{R}^d$, it holds that~$\abs{\nabla f_4(x) - \nabla f_4(y)}\leq 686L_1\abs{x-y}$.
\end{lemma}
\begin{proof}
In the following calculations, for any~$x\in\mathbb{R}^d\setminus\{0\}$, let~$\bar{y}$ denote the column vector~$\bar{y} = \abs{x}z - (z\cdot x) x/\abs{x}$. Note that~$\abs{\bar{y}}^2 = \abs{x}^2 - (z\cdot x)^2 \leq \abs{x}^2$. 
By direct calculation, 
it holds for any~$x\in\mathbb{R}^d\setminus\{0\}$ that
\begin{equation*}
\nabla f_4(x) = (1/2)(m_1-L_1)\partial H(z\cdot x/\abs{x}) \bar{y} + \big(m_1H(z\cdot x/\abs{x}) + L_1(1-H(z\cdot x/\abs{x}))\big) x
\end{equation*}
and
\begin{align*}
D^2 f_4(x) &= (1/2)(m_1-L_1)\partial^2H(z\cdot x/\abs{x})\bar{y}\bar{y}^{\top}/\abs{x}^2 \\
&\quad + \frac{(m_1-L_1)}{2} \partial H \bigg(\frac{z\cdot x}{\abs{x}}\bigg)\bigg(\frac{z x^{\top}}{\abs{x}} - \frac{ xz^{\top} + (x\cdot z) I_d}{\abs{x}} + \frac{(z\cdot x) x x^{\top}}{\abs{x}^3}\bigg)\\
&\quad+ (m_1-L_1) \partial H(z\cdot x/\abs{x}) x \bar{y}^{\top}/\abs{x}^2\\
&\quad+  \big(m_1H(z\cdot x/\abs{x}) + L_1(1-H(z\cdot x/\abs{x}))\big)I_d.
\end{align*}
It follows from~$\abs{\bar{y}}\leq \abs{x}$, the definition of~$\bar{y}$ and~$\abs{z} = 1$ that for any~$u,x\in\mathbb{R}^d\setminus\{0\}$ with~$\abs{u}=1$, we have~$\abs{D^2f_4(x)u}\leq  (1/2)L_1\max_{\mathbb{R}}\abs{\partial^2 H} + 2L_1\max_{\mathbb{R}}\abs{\partial H} + m_1+L_1$, which concludes the proof using~$\max_{\mathbb{R}}\abs{ \varphi} \leq 1$,~$\max_{\mathbb{R}}\abs{\partial \varphi} \leq 2$.
\end{proof}

Using arguments similar to the proof of Theorem~\ref{comp}, we have the following Theorem~\ref{comp2} for dissipative negative logdensities.

\begin{theorem}\label{comp2}
Assume the setting and notations of Theorem~\ref{comp} and~$d>2$. 
There exist bounded measurable~$\phi:\mathbb{R}^d\rightarrow\mathbb{R}$ and~$U\in C^1(\mathbb{R}^d)$ satisfying~$\nabla U(0)=0$ and for any~$x,y\in\mathbb{R}^d$ that
\begin{equation*}
x\cdot \nabla U(x)\geq m_1\abs{x}^2,\qquad \abs{\nabla U(x) - \nabla U(y)}\leq 686 L_1\abs{x-y}
\end{equation*}
such that if 
\begin{equation}\label{Nin3}
N < \frac{(\sin(2\cos^{-1}((\sqrt{3}+\sqrt{2})/4)))^{-d}}{3\abs{\hat{\theta}_N}(d+2)^{1/2}}-1
\end{equation}
then it holds that
\begin{equation}\label{compeq2}
\bigg|\mathbb{E}\bigg[\hat{\theta}_N(\hat{x}_N^{U,\phi}) - \int_{\mathbb{R}^d}\phi e^{-U}\bigg(\int_{\mathbb{R}^d} e^{-U}\bigg)^{-1}\bigg]\bigg|\geq \frac{\max_x\abs{\phi(x)}}{4}.
\end{equation}
\end{theorem}
\begin{remark}\label{cober2}
Note that~$U$ above satisfies the first condition in Corollary~1.6 in~\cite{MR2386063}, so that~$U$ satisfies a Poincar\'e inequality. On the other hand, by the results of~\cite{pmlr-v178-c}, existing sampling algorithms have complexity scaling linearly with the Poincar\'e constant. Theorem~\ref{comp2} thus implies that the Poincar\'e constant derived as a result of Corollary~1.6(1) in~\cite{MR2386063} scales exponentially with~$d$.
\end{remark}
\begin{proof}
Assume~\eqref{Nin3}. 
For any~$x\in\mathbb{R}^d$, let~$h_x:\mathbb{R}^d\rightarrow[0,1]$ be given by
\begin{equation*}
h_x(z) = \mathds{1}_{\{ y\neq 0: x\cdot y/\abs{y}> (\sqrt{3} + \sqrt{2})/4\}}(z).
\end{equation*}
For any~$z\in\mathbb{R}^d$, let~$f_z$ denote~$f_4$ as in~\eqref{f4def}. 
Suppose for contradiction that for any~$z\in\mathbb{R}^d$ with~$\abs{z}=1$, inequality~\eqref{compeq2} with~$U=f_z$ and~$\phi=h_z$ does not hold, in other words, it holds that
\begin{equation}\label{compeqneg2}
\bigg|\mathbb{E}\bigg[\hat{\theta}_N(\hat{x}_N^{f_z,h_z}) - \int_{\mathbb{R}^d}h_z e^{-f_z}\bigg(\int_{\mathbb{R}^d} e^{-f_z}\bigg)^{-1}\bigg]\bigg| < \frac{1}{4}.
\end{equation}
For any~$z$ with~$\abs{z}=1$, Lemma~\ref{f4mass} implies~$\int_{\mathbb{R}^d}h_z e^{-f_z}(\int_{\mathbb{R}^d}e^{-f_z} )^{-1} \geq \frac{1}{2}$. 
Therefore, it holds that~$\mathbb{E}[\hat{\theta}_N(\hat{x}_N^{f_z,h_z})] > 1/4$. 
If the event~$\mathcal{K}\subset\Omega$ defined by
\begin{equation*}
\mathcal{K}=\cap_{i=0}^N\{x_i^{f_z,h_z} \notin \{y\neq 0:z\cdot y/\abs{y}> (\sqrt{3}+\sqrt{2})/4 \}\}
\end{equation*}
satisfies the inequality~$\mathbb{P}(\mathcal{K}) \geq 1- 1/(4\abs{\hat{\theta}_N})$, then~\eqref{theb} implies~$\mathbb{E}[\hat{\theta}_N(\hat{x}_N^{f_z,h_z})] = \mathbb{E}[\mathds{1}_{\mathcal{K}}\hat{\theta}_N(\hat{x}_N^{f_z,h_z})] + \mathbb{E}[\mathds{1}_{\Omega\setminus\mathcal{K}}\hat{\theta}_N(\hat{x}_N^{f_z,h_z})] \leq 1/4$, 
which is a contradiction. Therefore~$\mathcal{K}$ satisfies~$\mathbb{P}(\mathcal{K})< 1-1/(4\abs{\hat{\theta}_N})$. 
Equivalently, it holds that
\begin{equation}\label{ppo2}
\mathbb{P}(A_z) \geq 1/(4\abs{\hat{\theta}_N}),
\end{equation}
where~$A_z := \{\exists i\in[0,N]\cap\mathbb{N}: x_i^{f_z,h_z}\in \{y\neq 0:z\cdot y/\abs{y}> (\sqrt{3} + \sqrt{2})/4\}\}$. 
Let~$\mathcal{G}\subset \mathbb{S}^{d-1}$ be a finite set such that for any~$\bar{x},\bar{y}\in\mathcal{G}$ with~$\bar{x}\neq\bar{y}$ it holds that~$\bar{x}\cdot\bar{y} < \cos(2\cos^{-1}((\sqrt{3}+\sqrt{2})/4))$ and such that it has size~$S(2\cos^{-1}((\sqrt{3}+\sqrt{2})/4))$, for~$S$ given for any~$\bar{\alpha}\in(0,\pi/2)$ by~\eqref{jga}. 
Similar to~\eqref{gsi}, we have
\begin{equation}
\abs{\mathcal{G}}\geq \sqrt{2\pi}(d+2)^{-1/2}(2\cos^{-1}((\sqrt{3}+\sqrt{2})/4))^{-1}(\sin(2\cos^{-1}((\sqrt{3}+\sqrt{2})/4)))^{2-d}.
\end{equation}
The elements in~$\mathcal{G}$ will be considered as values for~$z$ in the above. 
For any~$i\in[1,\abs{\mathcal{G}}]\cap\mathbb{N}$, let~$y_i\in\mathcal{G}$, such that~$y_i\neq y_j$ for~$i\neq j$, and 
let~$V_i=\{\bar{x}\neq 0:y_i\cdot\bar{x}/\abs{\bar{x}} > (\sqrt{3}+ \sqrt{2})/4\}$. 
By definitions, it holds for~$i\neq j$ that~$V_i\cap V_j = \emptyset$.
For any~$j\in[1,\abs{\mathcal{G}}]\cap\mathbb{N}$, define~$A_j:=A_{y_j}$.
By~\eqref{ppo2},~$A_j$ satisfies
\begin{equation}\label{Pa2}
\mathbb{P}(A_j) \geq 1/(4\abs{\hat{\theta}_N}) \qquad\forall j\in[1,\abs{\mathcal{G}}]\cap\mathbb{N}.
\end{equation}
On the other hand, for any~$M\in\mathbb{N}$ with~$M\in(N,\abs{\mathcal{G}})\neq\emptyset$ and~$j_0,\dots,j_M\in[1,\abs{\mathcal{G}}]\cap\mathbb{N}$ such that~$j_i\neq j_k$ for all~$i,k$, if~$\omega\in \cap_{i=0}^M A_{j_i} $ holds, then for any~$k\in[0,M]\cap\mathbb{N}$ there exists~$i\in[0,N]\cap\mathbb{N}$ satisfying~$x_i^{L_1\abs{\cdot}^2/2,0}(\omega) \in V_k$ by definition, which contradicts~$M>N$, with the consequence that~$\cap_{i=0}^M A_{j_i} = \emptyset$ holds. The rest of the proof follows in the same way as in the end of the proof for Theorem~\ref{comp}, in particular the arguments around~\eqref{enp} and~\eqref{enq}.
\end{proof}

\section{Sampling guarantees for the flattened distribution}\label{impl}
Although the results of Section~\ref{mse} show that, when given i.i.d. samples from the flattened proposal distribution~$\pi$, the bias and mean-square error of the IS estimator are well-controlled, it remains to be seen that at least approximate samples can be generated in polynomial time and that they suffice in place of i.i.d. samples. The purpose of this section is to demonstrate 
that implementable IS estimators using approximate samples enjoy quantitative guarantees with constants of polynomial scaling, under the assumption that~$U\in C^2$ satisfies~$\abs{D^2 U}\leq\bar{L}$ for some~$\bar{L}>0$ and that~$U$ is convex outside a ball.

In this section, 
the function~$T:[\min_{\mathbb{R}^d}U,\infty)\rightarrow[\min_{\mathbb{R}^d}U,\infty)$ from the introduction is fixed as
\begin{equation}\label{T0def}
T(y) = \begin{cases}
M+1 &\textrm{if }y\leq M,\\
M+1+(y-M)^3/4 - (y-M)^4/16 &\textrm{if } M<y<M+2,\\
y &\textrm{if }y\geq M+2.
\end{cases}
\end{equation}
In particular,~$T$ satisfies~\eqref{Tdef} with~$c=1$. 
Note that~$\nabla (T\circ U)$ can be evaluated using only evaluations of~$U$ and~$\nabla U$. 
This enables existing methodology on gradient-based MCMC samplers to be used for sampling~$\pi$ as in~\eqref{pidef}. 
This choice of~$T$ (and~$c$) is not claimed to be optimal. The purpose of this section is to provide theoretical proof that a concrete sampler with complete polynomial scaling exists under the conditions as mentioned in the introduction. Optimizing~$T$,~$c$,~$M$ (and other components of the IS procedure), as well as the resulting quantities in this section, is not addressed here.

The following Proposition~\ref{cob2} shows that if the original~$U$ is convex outside a ball and the value~$M$ in~\eqref{Tdef} is chosen appropriately, then the flattened proposal distribution~$\pi$ is weakly log-concave everywhere. 
Here, weakly logconcave means that the logdensity is twice differentiable with nonpositive definite Hessian everywhere. 
Moreover, if~$U$ has a bounded Hessian, then~$T\circ U$ also has Hessian that is bounded polynomially in the problem parameters. 
\begin{prop}\label{cob2}
Assume~$U\in C^2$. 
Let~$c_U,\mathcal{R}\geq 0$,~$L\geq m >0$ and~$x^*\in B_{\mathcal{R}}$ be such that~\eqref{A1eq} holds for all~$x\in\mathbb{R}^d$.
Suppose there 
exists~$\bar{L}_0,\bar{L}_1,\bar{L}_2>0$ such 
that
\begin{equation}\label{tps}
\abs{\nabla U(x)} \leq \bar{L}_0+\bar{L}_1\abs{x},\quad\abs{D^2U(x)u} \leq \bar{L}_{2} \abs{u},\quad u^{\top} D^2U(x) u \geq 0
\end{equation}
for all~$x\in\mathbb{R}^d\setminus B_{\mathcal{R}},u\in\mathbb{R}^d$. 
Let~$M = U(0) + c_U + 2L\mathcal{R}^2$. 
Let~$T:[\min_{\mathbb{R}^d}U,\infty)\rightarrow\mathbb{R}$ be 
given by~\eqref{T0def}. 
The Hessian of~$T\circ U$ 
satisfies~$D^2(T\circ U)(x)\geq 0$ and~$\abs{D^2(T\circ U)(x) u}\leq \hat{L}\abs{u}$ for all~$x,u\in\mathbb{R}^d$, where
\begin{equation}\label{Lhdef}
\hat{L} = \bar{L}_{2} + (3/4)\big(\bar{L}_0+\bar{L}_1\big(\mathcal{R}+\sqrt{m^{-1}(4+4c_U + 5L\mathcal{R}^2)}\,\big)\big)^2. 
\end{equation} 
Moreover, if instead of~\eqref{A1eq} and~\eqref{tps}, there exist~$L\geq m>0$ and~$R\geq 0$ such that~\eqref{set} holds and~$M=U(0)+LR^2/2$, then the Hessian of~$T\circ U$ satisfies the same inequalities as before with~$\hat{L}=L + 3L^2\cdot(R(1+L/m)+1/\sqrt{m})^2$.
\end{prop}
\begin{remark}
If in 
Proposition~\ref{cob2}, strong convexity of~$U$ outside~$B_{\mathcal{R}}$ is assumed, 
then Proposition~\ref{cob} yields~$c_U,\mathcal{R}\geq 0$,~$L\geq m >0$ and~$x^*\in B_{\mathcal{R}}$ such that~\eqref{A1eq} holds.
In this case the values in Proposition~\ref{cob2} need not be those inferred by Proposition~\ref{cob} (for example, the latter are always restricted to~$c_U=0$).
\end{remark}
\begin{proof}
By direct calculation, it holds for any~$x\in\mathbb{R}^d$ that
\begin{equation}\label{jkq}
D^2(T\circ U)(x) = \partial^2 T (U(x)) \nabla U(x) (\nabla U(x))^{\top} + \partial T(U(x)) D^2 U(x).
\end{equation}
In case of~\eqref{A1eq} and~\eqref{tps}, by~$M \geq \min_{\mathbb{R}^d}U + c_U + (L/2)\abs{2\mathcal{R}}^2 \geq \min_{\mathbb{R}^d}U + c_U + (L/2)\max_{x\in B_{\mathcal{R}}}\abs{x-x^*}^2 \geq \max_{B_{\mathcal{R}}} U$, it holds that~$B_{\mathcal{R}} \subset \{U\leq M\}$. In case of~\eqref{set}, the condition~$B_R \subset \{U\leq M\}$ also holds similarly. Hereafter let~$\bar{\mathcal{R}}\in\{\mathcal{R},R\}$. In both cases, it follows that for any~$x\in B_{\bar{\mathcal{R}}}$ that~$D^2(T\circ U)(x) = 0$ and for any~$x\in\mathbb{R}^d\setminus B_{\bar{\mathcal{R}}}$, the inequality~$D^2(T\circ U)(x)\geq 0$ holds by convexity of~$U$ and the assumptions on~$T$. 

For the upper bound, 
it follows by definition~\eqref{T0def} of~$T$ that the coefficients in~\eqref{jkq} may be bounded by~$\max_y\abs{\partial^2 T(y)} \leq \max_{z\in[0,2]}(3z/2 - 3z^2/4) = 3/4$ and~$\max_y \abs{\partial T(y)} \leq 1$ respectively. 
Moreover, in case of~\eqref{A1eq}, it holds that~$U(0) \leq \min_{\mathbb{R}^d} U + c_U + L\mathcal{R}^2/2$, so that~$M-\min_{\mathbb{R}^d}U\leq 2c_U + 5L\mathcal{R}^2/2$. 
Therefore by the left-hand inequality of~\eqref{A1eq}, the inequality~$U(x) > M+2$ holds for all
\begin{equation}\label{xrna}
x\in\mathbb{R}^d\setminus B_{\mathcal{R} + \sqrt{(2/m)(2+2c_U + 5L\mathcal{R}^2/2)}} \subset \mathbb{R}^d\setminus B_{\mathcal{R} + \sqrt{(2/m)(M+2-\min_{\mathbb{R}^d}U)}}.
\end{equation}
Since the support of~$\partial^2T$ is contained in~$[M,M+2]$, it holds that~$\partial^2 T(U(x)) =0$ for all~$x$ as in~\eqref{xrna}.
From~\eqref{jkq}, as well as the assumptions~\eqref{tps} valid on~$\{U>M\}\subset \mathbb{R}^d\setminus B_{\mathcal{R}}$, 
the first assertion then follows. In the case of~\eqref{set}, by Taylor's theorem it also holds for any~$x\in\mathbb{R}^d\setminus B_R$ that
\begin{align*}
U(x) &\geq U(Rx/\abs{x}) + (x-Rx/\abs{x})\cdot \nabla U(Rx/\abs{x}) + m\abs{x-Rx/\abs{x}}^2/2\\
&\geq U(0)-LR^2/2 -LR(\abs{x}-R) +m(\abs{x}-R)^2/2,
\end{align*}
so that~$U(x)\geq U(0)+LR^2/2+2=M+2$ for all
\begin{equation*}
x\in \mathbb{R}^d\setminus B_{2R(1+L/m) + 2/\sqrt{m}} \subset \mathbb{R}^d\setminus B_{R(1+L/m) + \sqrt{R^2L^2/m^2+2R^2L/m+4/m}}.
\end{equation*}
The second assertion follows by~$\partial^2T(U(x))=0$ for all~$\{U>M+2\}\supset\mathbb{R}^d\setminus B_{2R(1+L/m) + 2/\sqrt{m}}$ and~\eqref{jkq} as well as~$\abs{\nabla U(x)}\leq L\abs{x}$ for all~$x$ by~\eqref{set}.
\end{proof}

The main results of this section are the following Proposition~\ref{ldf} (cf.~\cite[Theorem~1.2]{MR3784496}) and its Corollary~\ref{ldf2}, 
see Remark~\ref{jjq} for a discussion of the consequences. 
\begin{prop}\label{ldf}
Assume the setting and notations in Proposition~\ref{cob2}. 
Let~$N\in\mathbb{N}\setminus\{0\}$ and~$\epsilon,\epsilon'\in(0,1]$. 
For any~$i\in[1,N]\cap\mathbb{N}$, let~$\bar{\pi}_i$ be a probability distribution on~$\mathbb{R}^d$ such that the total variation between~$\bar{\pi}_i$ and the probability distribution~$\pi$ given by~\eqref{pidef} is less than or equal to~$\epsilon$. 
Let~$(x^{(i)})_{i\in[1,N]\cap\mathbb{N}}$ be an independent sequence of~$\mathbb{R}^d$-valued r.v.'s such that~$x^{(i)}\sim \bar{\pi}_i$ for all~$i$. It holds that
\begin{equation}\label{cone}
\sup_{\phi}\mathbb{P}( \abs{\bar{\mu}^N(\phi) - \mu(\phi)} > \epsilon' ) \leq 1-(1-\epsilon)^N + \frac{4\rho}{(\epsilon')^2N},
\end{equation}
where~$\rho$ is given by~\eqref{rhob}, the supremum is over all bounded measurable~$\phi:\mathbb{R}^d\rightarrow\mathbb{R}$ with~$\sup_x\abs{\phi(x)}\leq 1$ and~$\bar{\mu}(\phi)$ is given by the right-hand side of~\eqref{introest} with~$x_i$ replaced by~$x^{(i)}$. 
In particular, for any~$\bar{\epsilon}\in(0,1]$, if~$N \geq 8\rho(\epsilon')^{-2}\bar{\epsilon}^{-1}$ and~$\epsilon \leq \bar{\epsilon}(4N)^{-1}$, then it holds that
\begin{equation}\label{cone2}
\textstyle \sup_{\phi}\mathbb{P}( \abs{\bar{\mu}^N(\phi) - \mu(\phi)} > \epsilon' ) \leq \bar{\epsilon}.
\end{equation}
\end{prop}
\begin{proof}
Let~$(u^{(i)})_{i\in[1,N]\cap\mathbb{N}}$ be an i.i.d. sequence of~$\mathbb{R}^d$-valued r.v.'s such that~$u^{(i)}\sim \pi$ for all~$i$. 
Let~$((\hat{x}^{(i)})_{i\in[1,N]\cap\mathbb{N}},(\hat{u}^{(i)})_{i\in[1,N]\cap\mathbb{N}})$ be a maximal coupling~\cite[Theorem~7.3]{MR1741181} of~$(x^{(i)})_{i\in[1,N]\cap\mathbb{N}},(u^{(i)})_{i\in[1,N]\cap\mathbb{N}}$. 
Let~$\phi$ be any function as in the assertion. 
Let~$\hat{\mu}^N(\phi),\mu^N(\phi)$ be given by~\eqref{introest} with~$x_i$ replaced by~$\hat{x}^{(i)}$ and~$\hat{u}^{(i)}$ respectively. 
If~$\abs{\hat{\mu}^N(\phi) - \mu(\phi)} >\epsilon'$, then at least one of the events~$\abs{\hat{\mu}^N(\phi) - \mu^N(\phi)} >0$,~$\abs{\mu^N(\phi) - \mu(\phi)} >\epsilon'$ must occur. Therefore 
it holds that
\begin{align}
\mathbb{P}(\abs{\bar{\mu}^N(\phi) - \mu(\phi)} >\epsilon' ) &=\mathbb{P}(\abs{\hat{\mu}^N(\phi) - \mu(\phi)} >\epsilon' )\nonumber\\
&\leq \mathbb{P}( \abs{\hat{\mu}^N(\phi) - \mu^N(\phi)} > 0) + \mathbb{P}(\abs{\mu^N(\phi) - \mu(\phi)}>\epsilon' ).\label{ndl}
\end{align}
By the assumption on the total variation between~$\bar{\pi}_i$ and~$\pi$, 
the first term on the right-hand side of~\eqref{ndl} satisfies~$\mathbb{P}( \abs{\hat{\mu}^N(\phi) - \mu^N(\phi)} > 0) \leq 1- (1-\epsilon)^N$. By Theorem~\ref{stu}, the second term on the right-hand side of~\eqref{ndl} satisfies
\begin{align*}
\mathbb{P}(\abs{\mu^N(\phi) - \mu(\phi)}>\epsilon' ) &\leq 
(\epsilon')^{-2}\mathbb{E}[\abs{\mu^N(\phi)-\mu(\phi)}^2] \leq 4\rho N^{-1}(\epsilon')^{-2},
\end{align*}
which implies~\eqref{cone}. For the last assertion, if~$\epsilon\leq \bar{\epsilon}(4N)^{-1}$, then~$\epsilon \leq 1- (1-\bar{\epsilon} (4N)^{-1}) \leq 1-e^{-\bar{\epsilon}/(2N)} \leq 1-(1-\bar{\epsilon}/2)^{\frac{1}{N}}$, so that~$1-(1-\epsilon)^N\leq \bar{\epsilon}/2$, which concludes given~$N\geq 8\rho (\epsilon')^{-2}\bar{\epsilon}^{-1}$ and~\eqref{cone}.
\end{proof}

Following~\cite[Corollary~1]{MR4700257}, we also have a bound on the bias and mean-square error. 
\begin{corollary}\label{ldf2}
Assume the setting and notations of Proposition~\ref{ldf}. For any~$\bar{\epsilon} \in (0,1]$, if~$N\geq 16\rho (\bar{\epsilon}\epsilon')^{-2}$ and~$\epsilon\leq \bar{\epsilon}(4N)^{-1}$, then it holds that
\begin{equation}\label{cone3}
\textstyle \sup_{\phi}\mathbb{E}[\abs{\bar{\mu}^N(\phi) - \mu(\phi)}] \leq 2\bar{\epsilon} + \epsilon',\qquad 
\textstyle \sup_{\phi}\mathbb{E}[(\bar{\mu}^N(\phi) - \mu(\phi))^2] \leq 4\bar{\epsilon} + (\epsilon')^2.
\end{equation}
\end{corollary}
\begin{remark}\label{jjq}
In the assumptions of Proposition~\ref{ldf} and Corollary~\ref{ldf2}, if~$c_U,\mathcal{R},L,m$ satisfy~\eqref{cocoa} for some~$\hat{c}\geq 1$, 
then Corollary~\ref{coco2} implies that~$\rho$ satisfies~\eqref{cocob} with~$c=1$. If for example we have in addition~$c_U=0$ or that~\eqref{cocoa} holds with~$c_U\neq0$ and~$\hat{c}=e^{1/(10e)}$, then~\eqref{cocob} with~$c=1$ is a universally constant bound on~$\rho$, see Remark~\ref{remo}\ref{crem}. 
In this case and under the assumptions of Proposition~\ref{cob2}, 
Proposition~\ref{ldf} and Corollary~\ref{ldf2} imply by Proposition~\ref{cob2} that any of the algorithms from~\cite[Section~7]{pmlr-v83-c},~\cite[Proposition~11]{pmlr-v195-f} and~\cite[Theorem~D.7]{MR4763254} 
achieves~\eqref{cone2} and~\eqref{cone3} with polynomial complexity in all parameters. Note that the latter two references give algorithms with randomized runtimes, so that the sense of complexity is in boundedness of runtimes in expectation or in probability. The moment estimates required (if any) for the guarantees in these references can be found for example in~\cite[Lemma~D.1]{chak2024r}. 
\end{remark}
\begin{proof}
For any measurable~$\phi$ with~$\sup_x\abs{\phi(x)}\leq 1$, the last assertion in Proposition~\ref{ldf} 
implies for any~$p\in\{1,2\}$ that
\begin{align*}
\mathbb{E}[\abs{\bar{\mu}^N(\phi) - \mu(\phi)}^p] &= \mathbb{E}[\abs{\bar{\mu}^N(\phi) - \mu(\phi)}^p\mathds{1}_{\abs{\bar{\mu}^N(\phi) - \mu(\phi)}>\epsilon'}] \\
&\quad+ \mathbb{E}[\abs{\bar{\mu}^N(\phi) - \mu(\phi)}^p\mathds{1}_{\abs{\bar{\mu}^N(\phi) - \mu(\phi)} \leq \epsilon'}]\\
&\leq 2^p\bar{\epsilon} + (\epsilon')^p,
\end{align*}
which is the assertion.
\end{proof}

\section{Examples}\label{examples}
In this section, examples of nonlogconcave target distributions from concrete cases are presented where the results of Section~\ref{mse} may be applied. 
In the case of Section~\ref{gauex}, conditions on Gaussian mixtures are given under which the approximate sampling results of Section~\ref{impl} may also be applied. In particular, under such conditions, polynomial sampling guarantees are shown to be theoretically attainable. 
\subsection{Gaussian mixtures}\label{gauex}
A principal example of multimodal measures is the Gaussian mixture, namely an (unnormalized) density of the form~$e^{-U(x)}\propto\sum_{i=1}^K a_ie^{-\frac{1}{2}(x-x_i)^{\top}S_i(x-x_i)}$ with~$K\in\mathbb{N}\setminus\{0\}$,~$a_i>0$,~$x_i\in\mathbb{R}^d$ and symmetric positive definite~$S_i\in\mathbb{R}^{d\times d}$ for all~$i$. 
These distributions can be infeasible to directly sample from if the number of modes~$K$ is exponentially large as a consequence of product structure~\cite{1211409,4217931}. 
In any case, sampling from a mixture with MCMC has in many cases not been treated directly. In particular, to the author's knowledge, no scalable guarantees have been provided for the unrestricted case, where the covariance matrices~$S_i^{-1}$ are permitted to vary for different~$i$. 
It is shown in this section that if the modes~$(x_i)_i$ are contained in a Euclidean ball with radius scaling at most like~$\sqrt{d}$, then the sampling problem is tractable using IS with tail-matching proposal. 
Note that outside this~$\sqrt{d}$ scaling, the counterexamples showing intractability are also closely related to Gaussian mixtures, as seen in Section~\ref{bober}.

It is worth mentioning that Gaussian mixtures with equal variance~($S_i=S_j$ for all~$i,j$) are special in that the sampling problem can be solved satisfactorily in polynomial time 
despite large distances between the modes~\cite{guo2024p,NEURIPS2018_c6ede20e,MR4700257,mou2019s,MR2521882} (see however~\cite{chehab2024p} for negative results using naive implementations). 
In the literature, the first manifestation of this difference between equal and unequal variances seems to be in the pair of papers~\cite{MR2521882,MR2495560}. 
Mixtures with equal variance, in dimensions larger than one, also have logdensities that are not necessarily strongly concave outside any Euclidean ball. However, Gaussian mixtures do satisfy the dissipativity condition~\eqref{disseq} in all cases, which is used below to obtain the mean-square error tractability.

The following Proposition~\ref{GauMix}, when combined with Proposition~\ref{diss}, implies that the logdensity~$U$ of the Gaussian mixture satisfies Assumption~\ref{A1}. Subsequently, Corollary~\ref{coco2} provides a quantitative condition on the parameters~$(x_i)_i,(S_i)_i$ under which the mean-square error of IS with tail-matching proposal is well-controlled. This is stated precisely in Corollary~\ref{GauMix2}. Finally, Proposition~\ref{sim} presents conditions for strong logconcavity outside a ball. 
\begin{prop}\label{GauMix}
Let~$K\in\mathbb{N}\setminus\{0\}$. For any~$i\in[1,K]\cap\mathbb{N}$, let~$a_i>0$ be such that~$\sum_ia_i=1$, 
let~$x_i\in\mathbb{R}^d$, let~$S_i\in\mathbb{R}^{d\times d}$ be a positive symmetric definite matrix and let~$m_i,L_i$ denote the smallest and largest eigenvalue of~$S_i$ respectively. Let~$U:\mathbb{R}^d\rightarrow\mathbb{R}$ be given by
\begin{equation}\label{ufd}
U=-\ln(\textstyle\sum_{i=1}^Ka_i\exp(-\frac{1}{2}(\cdot-x_i)^{\top}S_i(\cdot-x_i))).
\end{equation}
 Let~$R\geq 0$ be such that~$x_i\in B_R$ for all~$i$. The function~$U$ satisfies~\eqref{disseq} with~$\alpha=\min_i m_i/2$ and~$\beta=\max_i L_i R^2(L_i/m_i)/2$. Moreover,~$U$ satisfies the right-hand inequality in~\eqref{A1eq} with~$L = L_i$,~$c_U=1-\ln(a_i)$ and~$x^*=x_i$ for all~$i$.
\end{prop}
\begin{proof}
It holds for any~$x\in\mathbb{R}^d$ that
\begin{equation}\label{jdj}
\nabla U(x) = \bigg(\sum_{i=1}^K a_i S_i (x-x_i) e^{-\frac{1}{2}(x-x_i)^{\top}S_i(x-x_i)}\bigg)\bigg(\sum_{i=1}^K a_i e^{-\frac{1}{2}(x-x_i)^{\top}S_i(x-x_i)}\bigg)^{-1}.
\end{equation}
For each summand in the first bracket on the right-hand side, it holds by~$x_i\in B_R$ that
\begin{align*}
S_i(x-x_i)\cdot x & \geq m_i\abs{x}^2 - L_iR\abs{x}\\
& = m_i\abs{x}^2/2 + m_i(\abs{x}-L_iR/m_i)^2/2 - L_i^2R^2/(2m_i)\\
& \geq \min_jm_j\abs{x}^2/2 - \max_jL_j^2R^2/(2m_j).
\end{align*}
Together with~\eqref{jdj}, this implies the first assertion. For the second assertion, by definition, 
it holds for any~$x\in\mathbb{R}^d$ 
and~$i\in[1,K]\cap\mathbb{N}$ 
that
\begin{align*}
U(x) &\leq -\ln (a_i\exp(-(x-x_i)^{\top}S_i(x-x_i)/2))\\
&= -\ln(a_i)+(x-x_i)^{\top}S_i(x-x_i)/2\\
&\leq -\ln(a_i)+L_i\abs{x-x_i}^2/2
\end{align*}
and that~$U(x) \geq -\ln (\sum_j a_j) = 0$, which concludes.
\end{proof}
The (omitted) proof of the next Corollary~\ref{GauMix2} is a direct application of Proposition~\ref{GauMix}, Proposition~\ref{diss} and Corollary~\ref{coco2} with~$\hat{c}=1$. Quantitative variations of the following result also hold. For example, by considering~$\hat{c}\neq1$ in Corollary~\ref{coco2}, the weights~$(a_i)_i$ may be allowed to scale (inverse-polynomially) with~$d$, whilst keeping a good estimate on~$\rho$; see the first part of Remark~\ref{remo}\ref{crem}. 
\begin{corollary}\label{GauMix2}
Assume the setting and notations of Proposition~\ref{GauMix}. Denote~$\kappa = \max_j L_j/\min_jm_j$,~$L=\max_j L_j$ and~$a=\max_ja_j\in(0,1]$. Suppose 
it holds that
\begin{equation}\label{qcn}
10e\kappa \big(\sqrt{L} R\kappa + \kappa^{\frac{1}{2}}\big(8(1-\ln(a)) + 10 LR^2\kappa^2\big)^{\!\frac{1}{2}}\big)^2\leq d-1,
\end{equation}
then~$\rho,c$ with~$M$ all as given in Corollary~\ref{coco2} and Proposition~\ref{diss} satisfies
\begin{equation*}
\rho \leq (2e^c) \vee \bigg(\frac{2e a^{-2}(2+e^c)}{5^d\cdot\sqrt{d\pi/2}} + \frac{10ea^{-2} (1+e^c)}{(5/4)^{d/2}} \bigg).
\end{equation*}
\end{corollary}

Next, it is verified that the results of Section~\ref{impl} may also be applied. We make the simplification that the matrices~$S_i$ are of the form~$s_iI_d$, for some constants~$s_i >0$. Under the assumption that there exists a unique~$k$ with~$s_k = \min_i s_k$ (which is not a negligible assumption by the second paragraph of this Section~\ref{gauex}), Proposition~\ref{sim} makes explicit that the flattened mixture is logconcave with Lipschitz log-gradient everywhere. 
If a further quantitative condition (in the spirit of Corollary~\ref{GauMix2}) is assumed, 
then polynomial sampling guarantees for the original target mixture follow as in Remark~\ref{jjq}. 

\begin{prop}\label{sim}
Assume the setting and notations of Proposition~\ref{GauMix} with~$K\geq 2$. For any~$i\in[1,K]\cap\mathbb{N}$, let~$s_i>0$. Assume~$S_i = s_iI_d$ for all~$i$ and that there exists a unique~$k$ such that~$s_k = \min_i s_i$. Let~$m$ be such that~$s_m = \min_{i\neq k} s_i$ and denote~$\hat{s}=\max_i s_i$. The Hessian~$D^2U$ satisfies~$u^{\top}D^2 U(x) u \geq s_k\abs{u}^2/2$ and~$\abs{D^2U(x) u}\leq \hat{s}\abs{u}$ for all~$x\in \mathbb{R}^d\setminus B_{\mathcal{R}}$,~$u\in\mathbb{R}^d$, where~$\mathcal{R}=R^*\vee R\vee s^*\vee \sqrt{\beta/\alpha\,}$ with
\begin{align}
R^* &:= \frac{s_mR+s_k\abs{x_k}+\sqrt{((s_k\abs{x_k} + s_m R)^2 - (s_m - s_k)(s_mR^2-s_k\abs{x_k}^2-4\ln C))\vee 0}}{s_m - s_k},\label{rsdef}\\
C &:= 
8\hat{s}^2e^{\frac{s_k}{4}(s^* + \abs{x_k})^2 - \frac{s_m}{4}(s^*-R)^2}(s^* + R)^2 
/(a_ks_k),\nonumber\\
s^* &:= \big(s_k^2(\abs{x_k}+R)^2 + 4(s_m-s_k)[(s_k\abs{x_k}+s_mR)R+4]\big)^{\frac{1}{2}}(s_m-s_k)^{-1}.\label{ssdef}
\end{align}
Moreover, for any~$c_U,L$ as in Proposition~\ref{GauMix} and~$T$ as in~\eqref{T0def} with~$M=U(0)+c_U+2L\mathcal{R}^2$, 
it holds for any~$u,x\in\mathbb{R}^d$ that~$\abs{D^2 (T\circ U)(x)u}\leq \hat{L} \abs{u}$ and~$u^{\top}D^2(T\circ U)(x) u\geq 0$, 
where~$\hat{L}$ is given by~\eqref{Lhdef} with~$\bar{L}_0,\bar{L}_1,\bar{L}_2,m$ therein replaced by~$\hat{s}R,\hat{s},\hat{s},\alpha$ respectively.
\end{prop}
\begin{proof}
A direct calculation yields for any~$x\in\mathbb{R}^d$ 
that~$D^2U(x) 
= \bar{U}_1(x) + \bar{U}_2(x) + \bar{U}_3(x)$ with
\begin{align*}
\bar{U}_1 (x) &= 
e^{U(x)}
\big(\textstyle \sum_j s_ja_j e^{-\frac{s_j}{2}\abs{x-x_j}^2}\big)I_d,\\
\bar{U}_2(x) &= -
e^{U(x)}\big(\textstyle \sum_j a_j e^{-\frac{s_j}{2}\abs{x-x_j}^2} s_j^2(x-x_j)(x-x_j)^{\top}\big)\\
\bar{U}_3(x) &= e^{2U(x)}\big(\textstyle \sum_i a_i e^{-\frac{s_i}{2}\abs{x-x_i}^2}s_i(x-x_i)\big)\big(\textstyle \sum_j a_j e^{-\frac{s_j}{2}\abs{x-x_j}^2}s_j(x-x_j)\big)^{\top}.
\end{align*}
For~$\bar{U}_2,\bar{U}_3$, we control the sums by separating the contributions related to the~$k^{\textrm{th}}$ mixture component and the other components. For convenience, for any~$i\in[1,K]\cap\mathbb{N}$ let~$b_i:\mathbb{R}^d\rightarrow[0,\infty)$ be given by
\begin{equation*}
b_i(x) = a_i e^{-\frac{s_i}{2}\abs{x-x_i}^2}e^{U(x)}.
\end{equation*}
Note that by definition~\eqref{ufd} of~$U$, it holds that~$\sum_ib_i=1$ and
\begin{equation}\label{bjb}
b_i(x)\leq \frac{a_i e^{-\frac{s_m}{2}(\abs{x}-R)^2}}{a_ke^{-\frac{s_k}{2}(\abs{x}+\abs{x_k})^2}}\qquad 
\forall x\in\mathbb{R}^d\setminus B_R, i\neq k. 
\end{equation}
Let~$u\in\mathbb{R}^d$ satisfy~$\abs{u}=1$. 
It holds for any~$x\in\mathbb{R}^d$ that
\begin{align*}
u^{\top}\bar{U}_2(x)u &= - \textstyle \sum_{j=1}^K b_j(x)s_j^2((x-x_j)\cdot u)^2\\
&= - \textstyle \sum_{j=1}^K b_j(x)\big(s_j(x-x_j)\cdot u - s_k(x-x_k)\cdot u \big)^2 \\
&\quad- \textstyle \sum_{j=1}^K 2b_j(x)\big(s_j(x-x_j)\cdot u\big)\big(s_k(x-x_k)\cdot u \big) + \big(s_k(x-x_k)\cdot u\big)^2,
\end{align*}
so that adding~$\pm\big(\textstyle \sum_{j=1}^K b_j(x)\big(s_j(x-x_j)\cdot u\big)\big)^2$ on the right-hand side yields
\begin{align*}
&u^{\top}\bar{U}_2(x)u + \textstyle \sum_{j=1}^K b_j(x)\big(s_j(x-x_j)\cdot u - s_k(x-x_k)\cdot u \big)^2\\
&\quad\geq -\big(\textstyle \sum_{j=1}^K b_j(x)\big(s_j(x-x_j)\cdot u\big)\big)^2,
\end{align*}
the right-hand side of which is equal to~$-u^{\top}\bar{U}_3(x)u$. This implies
\begin{align}
u^{\top}(\bar{U}_2(x)+\bar{U}_3(x))u &\geq - \textstyle \sum_{j=1}^K b_j(x)\big(s_j(x-x_j)\cdot u - s_k(x-x_k)\cdot u \big)^2\nonumber\\
&= - \textstyle \sum_{j\neq k} b_j(x)\big(s_j(x-x_j)\cdot u - s_k(x-x_k)\cdot u \big)^2\nonumber\\
&\geq - \textstyle \sum_{j\neq k} 4b_j(x)\hat{s}^2(\abs{x}+R)^2.\label{bzb}
\end{align}
By~\eqref{bjb}, it holds for any~$j\neq k$ and~$x\in\mathbb{R}^d\setminus B_R$ that
\begin{align*}
b_j(x)(\abs{x}+R)^2 &\leq a_ja_k^{-1} e^{-\frac{s_m}{2}(\abs{x}-R)^2+\frac{s_k}{2}(\abs{x}+\abs{x_k})^2}(\abs{x}+R)^2,\\
&= a_ja_k^{-1} e^{-\frac{s_m}{4}(\abs{x}-R)^2+\frac{s_k}{4}(\abs{x}+\abs{x_k})^2}\big[e^{-\frac{s_m}{4}(\abs{x}-R)^2+\frac{s_k}{4}(\abs{x}+\abs{x_k})^2}(\abs{x}+R)^2\big],
\end{align*}
and by taking the derivative of the logarithm and finding the maximum in~$\abs{x}$, the expression in the square brackets on the right-hand side is maximized at
\begin{equation*}
\abs{x} = \frac{s_k\abs{x_k} + s_kR + \sqrt{s_k^2(\abs{x_k}+R)^2 + 4(s_m-s_k)((s_k\abs{x_k}+s_mR)R + 4)}}{2(s_m-s_k)} \leq s^*,
\end{equation*}
where~$s^*$ is given by~\eqref{ssdef}. 
Therefore, 
it holds for any~$j\neq k$ and~$x\in\mathbb{R}^d\setminus B_{R\vee s^*}$ that
\begin{equation*}
b_j(x)(\abs{x}+R)^2 \leq a_ja_k^{-1} e^{\frac{s_k}{4}(s^* + \abs{x_k})^2 - \frac{s_m}{4}(s^*-R)^2}(s^* + R)^2,
\end{equation*}
which may be substituted into~\eqref{bzb} to obtain
\begin{equation*}
u^{\top}D^2U(x)u \geq s_k - 
4\hat{s}^2a_k^{-1}e^{-\frac{s_m}{4}(\abs{x}-R)^2+\frac{s_k}{4}(\abs{x}+\abs{x_k})^2}\cdot e^{\frac{s_k}{4}(s^* + \abs{x_k})^2 - \frac{s_m}{4}(s^*-R)^2}(s^* + R)^2.
\end{equation*}
Solving for the first exponential on the right-hand side to be bounded above by~$C^{-1}$ (as given in the assertion), one obtains that such a bound is valid for~$\abs{x}\geq R^*$ (as in~\eqref{rsdef}). Thus by~$\max_ja_j\leq 1$ it holds for any~$x\in\mathbb{R}^d\setminus B_{R^*\vee R\vee s^*}$ that
\begin{equation*}
u^{\top}D^2U(x)u \geq s_k- 
\frac{4\hat{s}^2}{a_k}\cdot \frac{a_ks_k}{8\hat{s}^2}= s_k-
s_k/2 
=s_k/2.
\end{equation*}
as required.

For the upper bound on the Hessian, for any~$i$ let~$v_i:\mathbb{R}^d\rightarrow\mathbb{R}$ be given by~$v_i = s_i(\cdot-x_i)\cdot u$. For any~$x\in\mathbb{R}^d$ and by definition of~$\bar{U}_2,\bar{U}_3$, 
it holds that
\begin{equation*}
u^{\top}(\bar{U}_2 + \bar{U}_3)u = -\textstyle \sum_{j=1}^K b_jv_j^2 + (\sum_{j=1}^K b_jv_j)^2 
= -\textstyle \sum_{j=1}^K b_j (v_j-\sum_{j=1}^Kb_jv_j)^2 \leq 0,
\end{equation*}
so that~$u^{\top}D^2Uu\leq \hat{s}$. Together with the lower bound, this implies our bound on~$\abs{D^2U}$.

Lastly, by~\eqref{jdj}, it holds for any~$x\in\mathbb{R}^d$ that
\begin{equation*}
\abs{\nabla U(x)}\leq \hat{s}(\abs{x} + \textstyle \max_i\abs{x_i}) \leq \hat{s}(\abs{x} +R) 
\end{equation*}
\indent The last assertion follows by Propositions~\ref{GauMix},~\ref{diss} and~\ref{cob2}.
\end{proof}

\subsection{Bayesian neural networks}\label{bnn}
To demonstrate the generality of Assumption~\ref{A1} and condition~\eqref{cocoa}, 
we consider the Bayesian inference problem for multi-class classification 
using neural networks 
(see e.g.~\cite[Section~2
]{doi:10.1088/0}). 
Although Corollary~\ref{coco2} provides performance guarantees for IS given samples from the corresponding proposal~\eqref{pidef}, it is emphasized here that, in the following setting, whether the proposal distribution can be accurately sampled in polynomial time is not certain. 
\subsubsection{Mathematical setting}
Let~$K,I,p,\bar{L}\in\mathbb{N}\setminus\{0\}$, with~$I\geq 2$. The value~$K$ will be the number of samples,~$I$ will be the number of classes in the multi-class classification problem,~$p$ will be the number of features and~$\bar{L}$ will be the number of layers in the neural network. It will be shown in Proposition~\ref{bnnthm} that Assumption~\ref{A1} holds with parameters depending explicitly on~$K,I,p,\bar{L}$ (among other parameters such as the number of neurons in the layers). Let~$\{(x^{(i)},y^{(i)})\}_{i\in[1,K]\cap\mathbb{N}}$ be such that~$x^{(i)}\in\mathbb{R}^p$ and~$y^{(i)}\in [1,I]\cap\mathbb{N}$ for all~$i$. Each~$(x^{(i)},y^{(i)})$ represents a given data point, where~$x^{(i)}$ is a vector of~$p$ feature values and~$y^{(i)}$ represents the class to which the datapoint is known to belong. 
Let~$m_0=p$,~$m_{\bar{L}}=\hat{m}_{\bar{L}}=I$ 
and for any layer~$l\in[1,\bar{L}-1]\cap\mathbb{N}$, let~$m_l,\hat{m}_l\in\mathbb{N}\setminus\{0\}$. The value~$m_l$ denotes the number of neurons in the~$l^{\textrm{th}}$ layer post-activation and~$\hat{m}_l$ denotes the number of real-valued elements after the affine transformation at the~$l^{\textrm{th}}$ layer pre-activation. 
For any~$l\in[1,\bar{L}]\cap\mathbb{N}$, 
let~$\sigma^{(l)}:\mathbb{R}^{\hat{m}_l}\rightarrow\mathbb{R}^{m_l}$
be continuous functions outputting the neuron values at the~$l^{\textrm{th}}$ layer, 
with the final layer given by
\begin{equation}\label{sigL}
\sigma^{(\bar{L})}(\bar{x}) = (e^{\bar{x}_1}/\textstyle \sum_{j=1}^I e^{\bar{x}_j},\dots,e^{\bar{x}_I}/\textstyle \sum_{j=1}^I e^{\bar{x}_j})^{\top}
\end{equation}
for all~$\bar{x} = (\bar{x}_1,\dots,\bar{x}_I)\in\mathbb{R}^I$.

For any~$i\in[1,K]\cap\mathbb{N}$ and~$l\in[1,\bar{L}]\cap\mathbb{N}$, let~$z^{(l,i)}$
be column-vector-valued functions, given for any~$\bar{w}=(\bar{w}^{(l)})_{l\in[1,\bar{L}]\cap\mathbb{N}}$,~$\bar{b} = (\bar{b}^{(l)})_{l\in[1,\bar{L}]\cap\mathbb{N}}$, with matrix~$\bar{w}^{(l)}\in\mathbb{R}^{\hat{m}_l\times \sum_{k=0}^{l-1}m_k}$ and vector~$\bar{b}^{(l)}\in\mathbb{R}^{\hat{m}_l}$ for all~$l$, inductively by~$z^{(0,i)}=x^{(i)}$ and
\begin{equation}\label{zinduc}
z^{(l,i)} = z^{(l,i)}(\bar{w},\bar{b}) = \sigma^{(l)}(\bar{w}^{(l)} ((z^{(l-1,i)})^{\top},\dots,(z^{(0,i)})^{\top})^{\top} + \bar{b}^{(l)}).
\end{equation}
All elements of the weights~$(\bar{w}^{(l)},\bar{b}^{(l)})_l$ may form a vector of variables in the sampling state space~$\mathbb{R}^d$, with~$d=\sum_{l=1}^{\bar{L}} \hat{m}_l (\sum_{k=0}^{l-1}m_k + 1)$. Otherwise, in order to enforce network structure, a subset of these elements may be fixed, whilst subsets of the rest of the elements may be forced to be equal, with values to be inferred/trained. In this latter case,~$d\leq \sum_{l=1}^{\bar{L}} \hat{m}_l (\sum_{k=0}^{l-1}m_k + 1)$ and functions~$\hat{w}_1,\hat{w}_2$ will represent the structure. For example, in case of classical feedforward neural networks~\cite[Section~2
]{doi:10.1088/0}, in each layer~$l$, the (scalar) weights in~$\bar{w}^{(l)}$ corresponding to the layers~$l-2,l-3,\dots,1$ are fixed to~$0$, since each layer is only connected to the adjacent layers. In this case, we have~$d=\sum_{l=1}^{\bar{L}}\hat{m}_l(m_{l-1} + 1)$. More details are given in Section~\ref{fssub} and the main results in this section are also discussed explicitly for this case in Remark~\ref{ffnr}. 

\subsubsection{Flexibility and scope}\label{fssub}
In the main results of the section, assumptions will be made on the activations~$(\sigma^{(l)})_{l\in[1,\bar{L}-1]\cap\mathbb{N}}$ such that they are more or less restricted to be a component-wise sigmoidal activation~\cite[Section~6.3.2]{MR3617773}. It is only necessary to make this assumption on~$\sigma^{(l)}$ for layers~$l$ with direct connections to the final layer (including connections that have skipped layers in-between). The proofs follow verbatim in this case, but this weakening of the assumptions is not implemented so as to avoid complicating already quite technical statements. 
For layers~$l$ not directly connected to the final layer, there is freedom in choosing~$\sigma^{(l)}$. 
For example, they accommodate component-wise activations including rectified linear units (ReLU)~\cite[Section~6.3.1]{MR3617773}, layer normalizations~\cite{ba2016layernormalization} or softmax units~\cite[Section~6.2.2.3]{MR3617773}. 
In principle, transformers~\cite{NIPS2017_3f5ee243} fit into the setting here if the network is modified so that connections to the final layer are outputs of sigmoidal activations. 

Contrary to the assumptions to be made on~$\sigma^{(l)}$, sigmoidal activations are discouraged in~\cite[Section~6.3.2]{MR3617773} due to the vanishing gradient problem, where gradient-based learning becomes slow and inefficient because gradients of the activation functions have magnitude close to zero in a large part of their domains. 
However, it is worthwhile to note that gradients are forced to vanish in a region close to zero if samples from the tail-matching proposal are generated using gradient-based samplers. At the intuitive level, tail-matching instead learns the posterior by also evaluating the objective function itself, rather than just its gradient, 
whilst 
utilizing 
a blessing of dimensionality. 
In this sense, in justification for restrictions to sigmoidal activations, vanishing gradients are not necessarily inherently problematic for the Bayesian viewpoint.

Feedforward neural networks for multi-class classification~\cite[Section~2
]{doi:10.1088/0} may be recovered by setting~$\hat{m}_l=m_l=p$ for all~$l\in[0,\bar{L}-1]\cap\mathbb{N}$, then restricting the weights~$(\bar{w}^{(l)})_l$ to be of the form
\begin{equation*}
\bar{w}^{(l)} = \begin{pmatrix}
(\bar{w}^{(l,1)})^{\top}&0&\dots&0\\
 & \vdots& & \\
(\bar{w}^{(l,m_l)})^{\top}&0&\dots&0
\end{pmatrix},
\end{equation*}
where~$\bar{w}^{(l,k)}\in\mathbb{R}^p$ are column vectors for all~$k\in[1,m_l]\cap\mathbb{N}$, and finally setting~$\sigma^{(l)}:\mathbb{R}^{m_l}\rightarrow\mathbb{R}^{m_l}$ 
to be given for any~$l\in[1,\bar{L}-1]\cap\mathbb{N}$,~$\bar{x}=(\bar{x}_1,\dots,\bar{x}_p)\in\mathbb{R}^p$ by
\begin{equation*}
\sigma^{(l)}(\bar{x}) = 
 (g(\bar{x}_1),\dots,g(\bar{x}_p))^{\top}
\end{equation*}
where~$g:\mathbb{R}\rightarrow\mathbb{R}$ is a sigmoidal activation function (e.g.~$g=\tanh$).

Finally, it is worth mentioning that for example stochastic gradients~\cite{10.5555/3104482.3104568} are often used in Bayesian learning contexts. We forego related discussions and implementations beyond full sampling steps, as this is beyond the scope of the paper, but note that the condition given in Corollary~\ref{bnncor} is independent of the dataset size~$K$. 

\subsubsection{Scalable conditions for validity of importance sampling}

In the next Proposition~\ref{bnnthm}, the mathematical setting from above is assumed. The goal is to show that the loss function (see~\eqref{Udefbnn} below), with~$L^2$ regularization~\cite[Section~7.1.1]{MR3617773} (that is with Gaussian prior from the Bayesian viewpoint), 
satisfies Assumption~\ref{A1} and to relate the network architecture to the parameters in the assumption. 
Corollary~\ref{bnncor} then provides scalable conditions on the network under which IS is a feasible approach. See in particular Remark~\ref{ffnr} for the case of basic feedforward neural networks. 

In the statement of Proposition~\ref{bnnthm}, 
some technical definitions and assumptions will be made, which are intuitively simple. 
Beside the restrictions on~$\sigma^{(l)}$ already mentioned above, the main assumption is that the weights in the last layer are distinct variables in the sampling space. 
More specifically, the 
functions~$\hat{w}_1,\hat{w}_2$ 
will be introduced, 
as mentioned, to map the sampling variables to the (possibly multiple) positions where they appear in the network layers~\eqref{zinduc}. 
When introducing~$\hat{w}_1,\hat{w}_2$, the functions~$J_1,J_2$ will also be introduced, which map the index of a weight in the network (if it is not fixed to a constant) to the corresponding index of the sampling variable (which may be shared for different weights in the network layers). Whether a weight in the network is fixed to a constant is determined by the functions~$F_1,F_2$. 
The technical assumption is then made on~$J_1,J_2,F_1,F_2$. 
This technical condition is not stringent and it is satisfied in practical cases.
\begin{prop}\label{bnnthm}
For any~$l\in[0,\bar{L}]\cap\mathbb{N}$, let~$S_l=\sum_{r=1}^l\hat{m}_r\sum_{k=0}^{r-1}m_k$ and~$S_l'=\sum_{r=1}^l\hat{m}_r$. Let~$\alpha_1,\alpha_2,\beta>0$ and~$d_1\in\mathbb{N}\setminus\{0\}$,~$d_2\in\mathbb{N}$ be such that~$d_1+d_2=d$. 
Let~$J_1: [1,S_{\bar{L}}]\cap\mathbb{N}\rightarrow [1,d_1]\cap\mathbb{N}$,~$F_1:[1,S_{\bar{L}}]\cap\mathbb{N}\rightarrow\{0,1\}$ and~$F_2:[1,S_{\bar{L}}']\cap\mathbb{N}\rightarrow\{0,1\}$ be mappings and if~$d_2\neq 0$ then let~$J_2: [1,S_{\bar{L}}']\cap\mathbb{N}\rightarrow [1,d_2]\cap\mathbb{N}$ also be a mapping. 
For any~$j\in[1,S_{\bar{L}}]\cap\mathbb{N}$, let~$c_j\in\mathbb{R}$ and let~$\hat{c}=\max_k(1-F_2(k))c_k - \min_k(1-F_2(k))c_k$. 
Let~$\hat{w}_1:\mathbb{R}^{d_1}\rightarrow\mathbb{R}^{S_{\bar{L}}}$ and, if~$d_2\neq0$,~$\hat{w}_2:\mathbb{R}^{d_2}\rightarrow\mathbb{R}^{S_{\bar{L}}'}$ be functions 
satisfying for any~$i,j$ that~$(\hat{w}_i(\bar{x}))_j= F_i(j)\bar{x}_{J_i(j)} +(1-F_i(j))\mathds{1}_{\{2\}}(i)\,c_j$ 
for all~$\bar{x}=(\bar{x}_1,\dots,\bar{x}_{d_i})\in\mathbb{R}^{d_i}$. 
For any~$l\in[1,\bar{L}]\cap\mathbb{N}$, let~$\bar{w}^{(l)}:\mathbb{R}^{d_1}\rightarrow\mathbb{R}^{\hat{m}_l\times\sum_{k=0}^{l-1}m_k}$ and, if~$d_2\neq0$,~$\bar{b}^{(l)}:\mathbb{R}^{d_2}\rightarrow\mathbb{R}^{\hat{m}_l}$ be given for any~$\bar{x}\in\mathbb{R}^{d_1}$ and~$\bar{y}\in\mathbb{R}^{d_2}$ by
\begin{subequations}\label{wld}
\begin{align}
\bar{w}^{(l)}(\bar{x}) &= \begin{pmatrix}
(\hat{w}_1(\bar{x}))_{S_{l-1}+1} &\dots &(\hat{w}_1(\bar{x}))_{S_{l-1}+\sum_{k=0}^{l-1}m_k}\\
& \vdots & \\
(\hat{w}_1(\bar{x}))_{S_{l-1}+(\hat{m}_l-1)\sum_{k=0}^{l-1}m_k} &\dots &(\hat{w}_1(\bar{x}))_{S_{l-1}+\hat{m}_l\sum_{k=0}^{l-1}m_k}\\
\end{pmatrix},\\
\bar{b}^{(l)}(\bar{y}) & = ((\hat{w}_2(\bar{y}))_{S_{l-1}' + 1},\dots,(\hat{w}_2(\bar{y}))_{S_{l-1}' + \hat{m}_l })^{\top}.\label{wld2}
\end{align}
\end{subequations}
Denote~$\bar{w}=(\bar{w}^{(l)})_l$ and~$\bar{b}=(\bar{b}^{(l)})_l$. 
If the restriction of~$J_1
$ to~$F_1^{-1}(1)\cap(S_{\bar{L}-1},S_{\bar{L}}]
$ 
is injective and 
there exists a constant~$\sigma_{\max}>0$ such 
that~$|\sigma_j^{(l)}(\bar{x})| \vee \abs{x_r^{(i)}} \leq \sigma_{\max}$ for all~$\bar{x}\in\mathbb{R}^{\hat{m}_l}$,~$l\in[1,\bar{L}-1]\cap\mathbb{N}$,~$j\in[1,m_l]\cap\mathbb{N}$,~$i\in[1,K]\cap\mathbb{N}$ and~$r\in[1,p]\cap\mathbb{N}$, 
then the function~$U:\mathbb{R}^d\rightarrow[0,\infty)$, given for any~$\bar{x}\in\mathbb{R}^{d_1}$ and~$\bar{y}\in\mathbb{R}^{d_2}$ by
\begin{equation}\label{Udefbnn}
U(\bar{x},\bar{y}) =\alpha_1 \abs{\bar{x}}^2 + \alpha_2\abs{\bar{y}}^2 - \beta\sum_{i=1}^K \sum_{k=1}^I \mathds{1}_{\{k\}}(y^{(i)})\ln z_k^{(\bar{L},i)}(\bar{w}(\bar{x}),\bar{b}(\bar{y})),
\end{equation}
satisfies~\eqref{A1eq} 
with~$m=2(\alpha_1\wedge\alpha_2)$,~$L=4(\alpha_1\vee\alpha_2)$,~$x^* = 0$ and
\begin{align}
c_U&= \beta K(\hat{c}+\ln(I)) + \beta^2 K^2(\alpha_1\vee\alpha_2)^{-1}(m^*\sigma_{\max}^2+1), \nonumber \\
\mathcal{R} &= \sqrt{\beta K(\hat{c}+\ln I)/(\alpha_1\wedge\alpha_2)}
\label{rbn}
\end{align}
where~$m^* = \max_{j\in[1,\hat{m}_{\bar{L}}]\cap\mathbb{N}}\abs{F_1^{-1}(1)\cap (S_{\bar{L}-1}+(j-1)\textstyle \sum_{k=0}^{\bar{L}-1}m_k ,S_{\bar{L}-1}+j\textstyle \sum_{k=0}^{\bar{L}-1}m_k]}$.
\end{prop}
\begin{remark}\label{bnnrem}
\begin{enumerate}[label=(\roman*)]
\item \label{bnnrem1} In case of basic feedforward neural networks, the value~$m^*$ is equal to~$m_{\bar{L}-1}$, the number of neurons in the penultimate layer. Otherwise,~$m^*$ is bounded above by the total number of direct connections into the last layer, including those from previous layers which have skipped the layers in-between.
\item \label{bnnrem2} The choice to consider~$\alpha_1\neq\alpha_2$ is motivated by the difference in treatment of the weights and biases~\cite[Section~7.1]{MR3617773}. In fact, it is stated in this reference that regularizing the biases can be harmful. On the other hand, without regularization, the resulting distribution has infinite mass. 
One possibility is to consider training weights and biases separately, in an alternating manner and switching between sampling and optimization. 
Alternatively, the case where both weights and biases are sampled with~$\alpha_1=\alpha_2$ can still be relevant~\cite{CHANDRA2019315,doi:10.1088/0,TAMAMESRODERO2025100957}. Motivated by these considerations, in Corollary~\ref{bnncor}, the simplifying assumption~$\alpha_1=\alpha_2$ is made, which holds either by choice, or in the case where biases are assumed to be fixed from separate training processes. In the case where~$\alpha_1>\alpha_2$, the statement in Corollary~\ref{bnncor} holds with a cubic factor in~$\alpha_1/\alpha_2$ on the left-hand side of~\eqref{bnncoreq}.
\end{enumerate}
\end{remark}
\begin{proof}[Proof of Proposition~\ref{bnnthm}]
For any~$l\in[1,\bar{L}]\cap\mathbb{N}$,~$i\in[1,K]\cap\mathbb{N}$, let the column 
vector~$\hat{z}^{(l-1,i)}\in\mathbb{R}^{\sum_{k=0}^{l-1}m_k}$ 
denote~$\hat{z}^{(l-1,i)} = ((z^{(l-1,i)})^{\top},\dots,(z^{(0,i)})^{\top})^{\top}$,
so that~\eqref{zinduc} is
\begin{equation}\label{zinduc2}
z^{(l,i)} = \sigma^{(l)}(\bar{w}^{(l)} \hat{z}^{(l-1,i)} + \bar{b}^{(l)}).
\end{equation}
By the definition~\eqref{sigL} of~$\sigma^{(\bar{L})}$, it holds for any~$i\in[1,K]\cap\mathbb{N}$,~$k\in[1,I]\cap\mathbb{N}$ that
\begin{equation}\label{lnz}
-\ln (z_k^{(\bar{L},i)}) = \ln(\textstyle \sum_{j=1}^I e^{\hat{x}_j-\hat{x}_k}) \leq 
\ln I + \Big|\ln\Big(\max_{j,k} e^{\hat{x}_j - \hat{x}_k}\Big)\Big|,
\end{equation}
where~$\hat{x}=(\hat{x}_1,\dots,\hat{x}_I)\in\mathbb{R}^I$ denotes~$\hat{x} = \bar{w}^{(\bar{L})}\hat{z}^{(\bar{L}-1,i)}+\bar{b}^{(\bar{L})}$. 
Inequality~\eqref{lnz} implies
\begin{equation}\label{beq1}
-\ln(z_k^{(\bar{L},i)}) -\ln I
\leq \max_{j,k}\abs{\hat{x}_j-\hat{x}_k}\leq 2\max_j\abs{(\bar{w}^{(\bar{L})}\hat{z}^{(\bar{L}-1,i)})_j} + \max_{j,k}\abs{\bar{b}_j^{(\bar{L})}-\bar{b}_k^{(\bar{L})}}
\end{equation}
For the right-hand side, by the assumption on~$(\sigma^{(l)})_l$ and Cauchy-Schwarz, it holds for any~$j\in[1,I]\cap\mathbb{N}$,~$i\in[1,K]\cap\mathbb{N}$ that
\begin{equation*}
\abs{(\bar{w}^{(\bar{L})}\hat{z}^{(\bar{L}-1,i)})_j} 
\leq \Big|\textstyle \sum_{r=1}^{\sum_{k=0}^{\bar{L}-1}m_k}\bar{w}_{jr}^{(\bar{L})}\hat{z}_r^{(\bar{L}-1,i)}\Big| 
\leq \Big(\sum_{r=1}^{\sum_{k=0}^{\bar{L}-1}m_k} (\bar{w}_{jr}^{(\bar{L})})^2\Big)^{\frac{1}{2}} (m^*)^{\frac{1}{2}}\sigma_{\max},
\end{equation*}
Moreover, we may split the maximum over the~$\bar{b}^{(\bar{L})}$ terms in~\eqref{beq1} into the fixed (bounded by~$\hat{c}$) and variable (as in depending on the sampling variables) expressions, then use Young's inequality on the variable part to match the coefficient~$(m^*)^{1/2}\sigma_{\max}$. By definition of~$\hat{w}_1,\hat{w}_2$, this procedure yields for any~$\theta\in\mathbb{R}^d$ that 
\begin{equation}\label{xjb}
U(\theta) \leq (\alpha_1\vee\alpha_2)\abs{\theta}^2 + \beta K\ln I +  2\beta K(m^*\sigma_{\max}^2+1)^{1/2}\abs{\theta} + \beta K\hat{c}.
\end{equation}
Using Young's inequality again, we obtain
\begin{equation*}
2(m^*\sigma_{\max}^2+1)^{1/2}\abs{\theta} \leq (\alpha_1\vee\alpha_2)\abs{\theta}^2/(\beta K) + \beta K(\alpha_1\vee\alpha_2)^{-1}(m^*\sigma_{\max}^2+ 1),
\end{equation*}
so that
\begin{equation*}
U(\theta) \leq 2(\alpha_1\vee\alpha_2)\abs{\theta}^2 + \beta K\ln I + \beta K\hat{c} + \beta^2 K^2(\alpha_1\vee\alpha_2)^{-1}(m^*\sigma_{\max}^2+1),
\end{equation*}
which is the right-hand inequality in~\eqref{A1eq} with~$c_U,L,x^*$ as in the assertion.\\
\indent Now, by definition~\eqref{sigL} of~$\sigma^{(\bar{L})}$ and the assumption on~$\hat{w}_1,\hat{w}_2$, it holds that~$U\geq (\alpha_1\wedge \alpha_2)\abs{\cdot}^2$ and~$U(0) \leq \beta K(\hat{c} + \ln I)$. 
For~$\mathcal{R}$ given by~\eqref{rbn}, this implies for any~$(\bar{x},\bar{y})\in\mathbb{R}^d\setminus B_{\mathcal{R}}$ that
\begin{align*}
U(\bar{x},\bar{y}) &\geq (\alpha_1\wedge\alpha_2)\abs{(\bar{x},\bar{y})}^2 \\
&\geq \min_{\mathbb{R}^d}U +(\alpha_1\wedge\alpha_2)(\abs{(\bar{x},\bar{y})}^{2}-\mathcal{R}^{2})\\
&\geq \min_{\mathbb{R}^d}U +(\alpha_1\wedge\alpha_2)(\abs{(\bar{x},\bar{y})}-\mathcal{R})^2,
\end{align*}
which is the left-hand inequality in~\eqref{A1eq}. 
\end{proof}
In the following Corollary~\ref{bnncor}, an explicit condition on the number of classes~$I$ in the classification problem, the number~$m^*$ (which is at most the number of neurons directly connected to the final network layer, see Remark~\ref{bnnrem}\ref{bnnrem1}), the maximum value~$\sigma_{\max}$ of the activation functions directly inputted into the final layer, the regularization weights~$\alpha_1=\alpha_2$ (see Remark~\ref{bnnrem}\ref{bnnrem2}), the maximum difference~$\hat{c}$ between fixed bias values and the number of sampling variables~$d$ is provided under which tail-matching IS produces accurate estimates given samples from the proposal distribution. 
\begin{corollary}\label{bnncor}
Assume all of the presuppositions and notations in Proposition~\ref{bnnthm}. Assume moreover~$\beta=1/K$ and~$\alpha_1=\alpha_2$. 
If
\begin{equation}
40e \big(2(\hat{c}+\ln I)^{1/2} + \alpha_1^{-\frac{1}{2}}(m^*\sigma_{\max}^2 + 1)^{\frac{1}{2}} \big)^2 \leq d-1,\label{bnncoreq}
\end{equation}
then~$\rho,c$ as in Theorem~\ref{stu} with~$M=
U(0)$ satisfies
\begin{equation*}
\rho \leq (2e^c) \vee \bigg(\frac{2(2+e^c)}{e\sqrt{d\pi/2}} + \frac{1+e^c}{(1-\sqrt{4/5})e}\bigg).
\end{equation*}
\end{corollary}
\begin{remark}\label{ffnr}
We consider when condition~\eqref{bnncoreq} is satisfied in classical feedforward networks~\cite[Section~2
]{doi:10.1088/0} with layers of the same width (except the first layer of width~$p$ and the last layer of width~$I$). 
In this setting, we have~$m^*=m_{\bar{L}-1}$ (see Remark~\ref{bnnrem}\ref{bnnrem1}). 
In addition, 
we may set~$\hat{c}=0$ (since no weights or biases are fixed) 
and it holds that~$d = m_1(p+1) + m_{\bar{L}-1}(m_{\bar{L}-1}+1)(\bar{L}-2) + I(m_{\bar{L}-1} + 1) > m_{\bar{L}-1}(m_{\bar{L}-1}+1)(\bar{L}-2)$ (which is the number of weights and biases in each intermediate layer multiplied by the number of intermediate layers in the network). 
If we neglect the logarithmic and constant (in~$m^*$) terms and assume~$\sigma_{\max}=1$, then condition~\eqref{bnncoreq} is satisfied when the total number of (intermediate) neurons~$m_{\bar{L}-1}(\bar{L}-2)$ is larger than~$109\alpha_1^{-1}$, where~$\alpha_1$ is the~$L^2$ regularization weight. 
\end{remark}
\begin{proof}
Recall the definitions of~$m,L,c_U,\mathcal{R}$ given in Proposition~\ref{bnnthm}. 
The quantity~$\bar{R}$ given by~\eqref{Rbdef} satisfies
\begin{equation}\label{srf}
\bar{R} = \alpha_1^{-1/2}(\hat{c}+\ln I)^{1/2} + \alpha_1^{-1/2}\Big(U(0)-\min_{\mathbb{R}^d}U\Big)^{1/2}
\end{equation}
We estimate the expression in the last bracket. By definition~\eqref{Udefbnn} of~$U$ and the assumption~\eqref{sigL} on the final output layer (essentially the first equation in~\eqref{lnz}), it holds for any~$\theta\in \mathbb{R}^d$ that
\begin{equation*}
U(\theta) = \alpha_1\abs{\theta}^2 + K^{-1}\textstyle \sum_{i=1}^K\sum_{k=1}^I\mathds{1}_{\{k\}}(y^{(i)}) (\ln \sum_{j=1}^I e^{\hat{x}_j} - \hat{x}_k),
\end{equation*}
where~$\hat{x}=(\hat{x}_1,\dots,\hat{x}_I)$ is the same as in the proof of Proposition~\ref{bnnthm}. The summands (as functions of~$\hat{x}$) have gradients~$e^{\hat{x}}/\abs{e^{\hat{x}}}_1 - e_k$ for~$k=y^{(i)}$ and zero otherwise, where~$\abs{\cdot}_1$ is the~$l^1$-norm and~$e_k$ denotes the~$k^{\textrm{th}}$ Euclidean basis vector. 
The~$l^1$-norm of this gradient is clearly bounded above by~$2$. 
Therefore by (the second inequality in)~\eqref{beq1} and the same arguments leading to~\eqref{xjb}, 
it holds for any~$\theta\in\mathbb{R}^d$ that
\begin{equation*}
\abs{U(\theta) - \alpha_1\abs{\theta}^2 - U(0)} 
\leq 2(m^*\sigma_{\max}^2+1)^{1/2}\abs{\theta} + \hat{c},
\end{equation*}
so that
\begin{align*}
U(\theta) &\geq \alpha_1\abs{\theta}^2+U(0)-2(m^*\sigma_{\max}^2+1)^{1/2}\abs{\theta} -\hat{c} \\
&= U(0) + (\alpha_1^{1/2}\abs{\theta} - (m^*\sigma_{\max}^2+1)^{1/2}\alpha_1^{-1/2})^2 - (m^*\sigma_{\max}^2+1)\alpha_1^{-1}-\hat{c},
\end{align*}
and by taking minima on both sides
\begin{equation*}
\textstyle\min_{\mathbb{R}^d}U(\theta)\geq U(0)- (m^*\sigma_{\max}^2+1)\alpha_1^{-1}-\hat{c}.
\end{equation*}
Substituting into~\eqref{srf} yields
\begin{align*}
\bar{R} &\leq \alpha_1^{-1/2}(\hat{c}+\ln I)^{1/2} + \alpha_1^{-1}(m^*\sigma_{\max}^2+1)^{1/2} + \alpha_1^{-1/2}\hat{c}^{1/2}\\
&\leq 2\alpha_1^{-1/2}(\hat{c}+\ln I)^{1/2} + \alpha_1^{-1}(m^*\sigma_{\max}^2+1)^{1/2}.
\end{align*}
Therefore, 
denoting~$\kappa=L/m=2$, 
it holds by~\eqref{bnncoreq} that
\begin{equation*}
L\bar{R}^2\kappa \leq 8(2(\hat{c}+\ln I)^{1/2} + \alpha_1^{-1/2}(m^*\sigma_{\max}^2+1)^{1/2})^2.
\end{equation*}
Thus substituting into~\eqref{jjv} and using~\eqref{bnncoreq} 
to bound~$e^{2c_U}\leq e^{(d-1)/(20e)}$ in the same expression concludes the proof.
\end{proof}

\paragraph{Acknowledgments}
The author thanks Giacomo Zanella for feedback and discussions. Support from the European Research Council (ERC),
through StG “PrSc-HDBayLe” grant ID 101076564, is acknowledged.

\bibliography{document}       

\end{document}